\documentclass[10pt,twocolumn,twoside]{IEEEtran}
\usepackage{cite}
\usepackage{bm}
\usepackage{stfloats}
\usepackage{amsmath,amssymb,amsfonts}
\usepackage{graphicx}
\usepackage{textcomp}
\usepackage{xcolor}
\usepackage{times}
\usepackage{subfigure}
\usepackage{amssymb,amsmath}
\usepackage{siunitx}
\usepackage{algorithm}
\usepackage{balance}
\usepackage{algorithmic}
\usepackage{caption}
\usepackage{hyperref} %
\hypersetup{hypertex = true,
	colorlinks = true, %
	linkcolor = black, %
	urlcolor = blue,%
	citecolor = blue} %

\def\BibTeX{{\rm B\kern-.05em{\sc i\kern-.025em b}\kern-.08em
		T\kern-.1667em\lower.7ex\hbox{E}\kern-.125emX}}

\newcommand{\mb}[1]{{  \mathbf  #1}}  
\newtheorem{lemma}{\bf Lemma}
\newtheorem{corollary}{\bf Corollary}

\begin{document}
	
	\bibliographystyle{IEEEtran}
	
	\title{Near-Field Wideband Beamforming for Extremely Large Antenna Arrays}
	\author{\IEEEauthorblockN{Mingyao Cui,~\IEEEmembership{Graduate Student 
	Member,~IEEE}, Linglong Dai,~\IEEEmembership{Fellow,~IEEE}
}
\thanks{All authors are with the Department of Electronic Engineering, 
Tsinghua University, Beijing 100084, China (e-mails: cui-my16@tsinghua.org.cn, 
daill@tsinghua.edu.cn).}
\thanks{This work was funded in part by the National Science Fund for 
Distinguished Young Scholars (Grant No. 62325106), and in part by the National 
Key R\&D Program of 
	China 
(No. 2023YFB3811503). \emph{(Corresponding author: Linglong Dai.)} }
}

	\maketitle
	\IEEEpeerreviewmaketitle
	
	\begin{abstract}

The natural integration of extremely large antenna arrays (ELAAs) and terahertz 
(THz) communications can 
potentially establish Tbps data links for 6G
networks. 
However, due to the extremely large array aperture and wide 
bandwidth, a new phenomenon termed as ``near-field beam split" emerges. This 
phenomenon causes beams at different frequencies to focus on 
distinct physical locations, leading to a significant loss of 
the beamforming gain.   
To address this challenging problem, we first harness a piecewise-far-field 
channel model to approximate the complicated near-field wideband channel. In 
this model, the entire large array is partitioned into several small 
sub-arrays. While the wireless channel's phase discrepancy across the entire 
array is modeled as near-field spherical, the phase discrepancy within each 
sub-array is approximated as far-field planar.
Built on this approximation, a phase-delay focusing (PDF) method employing 
delay phase precoding (DPP) architecture is proposed. Our PDF method could 
compensate for the 
intra-array 
far-field phase discrepancy and the inter-array near-field phase discrepancy 
via the joint control of phase shifters and time delayers, respectively. 
Theoretical and numerical 
results are provided to demonstrate the efficiency of the proposed PDF method 
in mitigating the near-field beam split effect.
Finally, we define and derive a novel metric termed as the ``effective 
Rayleigh 
distance" by the evaluation of beamforming gain loss. Compared to classical 
Rayleigh distance, the effective Rayleigh distance is more accurate in 
determining the near-field range for practical communications.

%

	\end{abstract}
	
	\begin{IEEEkeywords}
		Extremely large antenna array, wideband, near-field beam split, 
		beamforming, Rayleigh distance.
	\end{IEEEkeywords}
	
	\section{Introduction}\label{sec:1}
	As a key technology for 5G communication systems, large antenna arrays 
	(LAAs) could improve the transmission rate by orders of magnitude via 
	efficient beamforming/precoding \cite{Overview_Heath2016}. To further reap 
	the 
	benefits of massive antennas, LAAs are evolving 
	to extremely 
	large antenna arrays (ELAAs) for 6G communications  
	\cite{XLMIMO_Carvalho2020}, 
	where the array aperture is dramatically increased to support 
	ultra-high-speed communications. There are abundant possible 
	implementations of ELAA. For instance, 
	ELAA could be employed in distributed 
	multiple-input-multiple-output (MIMO) systems relying on radio 
	stripes \cite{radiostrip_Ericsson19} or in reconfigurable intelligent 
	surface (RIS) systems \cite{RISNF_Tang2021} to improve the 
	network capacity. It is also envisioned to coat ELAAs on 
	entire walls to enhance the coverage of wireless signals 
	\cite{IE_Nie19}.

	In addition, ELAAs are usually combined with high-frequency communications. 
	Benefiting from the abundant spectrum 
	resources, terahertz (THz) communications can 
	provide a very large bandwidth of several GHz, allowing for Tbps data rates 
	for 6G
	networks\cite{6Gchallenge_Rappaport2019}. Besides, the extremely small size
	of THz antennas also favorably facilitates the deployment of ELAAs 
	\cite{THzsurvey_Elayan2020}. 
	As a consequence, the natural integration of THz wideband 
	communications and 	ELAAs has been regarded as a 
	pivotal candidate for  next-generation wireless networks 
	\cite{THzsurvey_Elayan2020}.

	\subsection{Prior Works}
	The evolution from LAA to ELAA not only implies 
	a sharp increment in array aperture, but also leads to a fundamental change 
	in 
	the characteristics of the electromagnetic (EM) field \cite{NearMag_Cui23}. 
	The  electromagnetic radiation field can generally be divided into the 
	far-field and radiation near-field regions \cite{NearLoS_Zhou2015,  
	NearCE_Fried2019}. In the far-field region, the wireless channel could be  
	modeled 
	under the \textbf{planar} wave assumption, where the phase of the array 
	response vector is a \emph{linear} function of the antenna index 
	\cite{FundWC_Tse2015,ModularMIMO_2023}. 
	In contrast, the wavefront of near-field channel has to be modeled 
	accurately as \textbf{spherical}, where the phase of near-field array 
	response vector is a 
	\emph{non-linear} function of the antenna index 
	\cite{NearLoS_Zhou2015}.
 The boundary between near-field and far-field is typically 
	quantified by 
	the Rayleigh distance \cite{NearLoS_Zhou2015}, also known as the 
	Fraunhofer distance \cite{fresnel_Selvan2017}, which is proportional to 
	the square of array aperture normalized by wavelength. 
	Since the array aperture is typically not very large in the current 5G 
	systems, the near-field range of 5G LAA is negligible. That is why 
	classical beamforming techniques usually direct a beam with planar 
	wavefront in a specific direction \cite{Overview_Heath2016}. 
	In contrast, as the number of ELAA's antennas increases dramatically, the 
	Rayleigh distance will be expanded by orders of magnitude. 
	The near-field range of an ELAA could be up to several hundreds 
	of meters \cite{NearMag_Cui23}, covering a large part of typical cells. 
	In this scenario, it is necessary to perform near-field beamforming to 
	focus 
	the energy of a beam on a desired user location \cite{NearMag_Zhang23} 
	by exploiting the spherical wavefront property. 
	Given this non-negligible near-field range, near-field communications will be of pivotal significance in next-generation communications.
	
	 Moreover, when it comes to wideband systems, another critical 
	change in 
	EM waves known as beam split is induced, which also 
	has the terminology ``beam squint"~\cite{THzOverview_Sar2021, 
		BSprecoding_Hanzo2020}.
	Classically, LAA relies on phased shifter (PS) based analog 
	beamforming architecture, allowing only  
	\emph{frequency-independent} phase shifting for narrowband 
	beamforming \cite{precoding_Sohrabi2017}. 
	However, the array response vectors of wireless channels are 
	\emph{frequency-dependent}, especially for THz wideband networks, causing 
	the 
	wavefront of a beam at different frequencies to deviate from that at the 
	center 
	frequency. 
	To elaborate, in far-field scenarios, the beam split effect makes 
	beams of different frequency components propagate in distinct 
	directions  \cite{THzOverview_Sar2021}. 
	On the other hand, in near-field scenarios, the beam split effect results 
	in 
	a new phenomenon where beams at different frequencies are focused at 
	varying directions and distances. Consequently, the signal energy 
	fails to converge on the desired receiver's location. 
	Only signals around the center frequency can be captured 
	by the receiver, while most beams with frequencies far away from the 
	center frequency suffer from an unacceptable beamforming gain loss.

	Over the past few years, intensive research has been devoted to studying 
	advanced beamforming technologies to address the 
	far-field beam split effect \cite{BSprecoding_Hanzo2020, 
		STBC_Liu2019,MOALT_Yu2016, TTD_Lin2017, DPP_Tan2019, DPP_Liao21
		}.  
	Relevant methods fall into two primary
	categories, i.e., algorithmic methods and hardware-based mitigation 
	methods. In the first category, researchers have endeavored to generate 
	wide beams by carefully optimizing the PSs to achieve 
	flattened beamforming gain across the entire bandwidth
	\cite{BSprecoding_Hanzo2020, 
	STBC_Liu2019,MOALT_Yu2016}. While these algorithms are relatively 
	straightforward to implement in practice, their beamforming performance is 
	severely hindered by the 
	presence of beam split as well. This limitation arises because they still 
	rely on PS-based analog beamforming. 
	The second category of solutions employs true-time-delay (TTD) circuits  
	instead of 
	PSs to generate \emph{frequency-dependent} beams, which offers the 
	potential to eliminate far-field beam split 
	\cite{TTD_Lin2017,DPP_Tan2019, DPP_Liao21 
	}. Inspired by this idea, several array structures, such as true-time-delay 
	(TTD) arrays \cite{TTD_Lin2017} and
	delay-phase precoding (DPP) arrays \cite{DPP_Tan2019,DPP_Liao21},  
	have been envisioned and developed to counteract the far-field beam split 
	effect.
	Despite the rapid development of solutions to far-field beam split, it is 
	essential to note that the aforementioned methods are 
	all customized for far-field communications. They are not  
	applicable to tackle the challenges posed by near-field beam split, because
	the models of far-field channels and near-field channels differ remarkably. 
	To 
	the best of our knowledge,  the  near-field beam split effect has not been 
	studied in the literature.
	
\subsection{Our Contributions}
	To fill in this gap, a  phase-delay focusing (PDF) method is proposed to 
	tackle the near-field beam split problem\footnote{Simulation codes are 
	provided to reproduce the results presented in this article:  
	http://oa.ee.tsinghua.edu.cn/dailinglong/publications/publications.html.}. 
	Our key 
	contributions are
	summarized as follows.
	\begin{itemize}
		\item 	First, we introduce the  near-field beam split effect of ELAA 
		by 
		comparing the loss of beamforming gain resulting from both far-field 
		and near-field beam split effects. We formulate the model of near-field 
		beam 
		split effect and reveal that this
		effect causes beams at different frequencies to focus on distinct 
		locations.  
		\item 	Second, a piecewise-far-field wideband channel model is 
		proposed 
		to approximate the near-field wideband channel model with high 
		accuracy. 
		In this model, the entire ELAA is partitioned into multiple small 
		sub-arrays. In this way, we could reasonably assume that the receiver 
		is located in the 
		far-field region 
		of each small sub-array while being in the 
		near-field region of the entire ELAA. This partition allows us to 
		decompose the 
		complicated phase discrepancy of a near-field channel into two distinct 
		components: 
		the 
		inter-array near-field phase discrepancy and the intra-array far-field 
		phase discrepancy.
		Leveraging this decomposition, a phase-delay focusing (PDF) method is 
		proposed based on the DPP array architecture, where the inter-array 
		phase and the intra-array phase are compensated by the PSs and TTDs of 
		DPP, respectively. Simulation results validate the efficacy of the 
		proposed PDF method in mitigating the near-field beam split effect.
		\item Finally, by evaluating the gain loss of far-field beamforming in 
		the near-field region, a new 
		metric called effective Rayleigh distance is defined to 
		distinguish the far-field and near-field regions. 
		Classical Rayleigh distance, which is defined by evaluating the phase 
		error between planar wave and spherical wave, is not precise enough to 
		capture the near-field region where far-field beamforming methods are 
		not applicable. To tackle this problem, we conduct a theoretical 
		evaluation on the 
		gain 
		loss of far-field beamforming in the physical space. Subsequently,  the 
		close-form expression of effective Rayleigh distance is derived, which 
		defines the region where the gain loss of far-field beamforming exceeds 
		a threshold.
        Since beamforming gain directly affects the received signal power, our
		proposed effective Rayleigh distance is a more accurate metric 
		for measuring the 
		near-field range for practical communications.  
	\end{itemize}

	\subsection{Organization and notation}\label{sec:1.3}
	The rest of the paper is organized as follows. In section \ref{sec:2}, the 
	wideband ELAA channel model is introduced and the near-field beam split 
	effect is discussed. In section 
	\ref{sec:3}, the proposed piecewise-far-field channel 
	model and the proposed PDF method are explained in detail. Theoretical 
	analysis on the beamforming gain of the PDF method is also offered.
	Section \ref{sec:4} elaborates on the effective Rayleigh distance. 
	Numerical results are provided in section \ref{sec:5}. Finally, conclusions 
	are drawn in section 
	\ref{sec:6}.
	
	\emph{Notation}: 
	Lower-case boldface letters $\mb{x}$  denote vectors; 
	$(\cdot)^T$, $(\cdot)^H$, $(\cdot)^*$ and $\|\cdot\|_k$ denote the 
	transpose, conjugate transpose, conjugate,
	and $k$-norm of a vector or matrix respectively; $|x|$ denotes the the 
	amplitude of scalar $x$; $\arg(x)$ denotes the phase of $x$; $[\mb{x}]_n$ 
	represents the $n^{\rm th}$h element of vector
	$\mb{x}$; $[\mb{X}]_{ij}$ 
	represents the $(i,j)^{\rm th}$ entry of matrix
	$\mb{X}$; 
	$\mathcal{CN}(\mu; \Sigma)$ and $\mathcal{U}(a; b)$ denote the Gaussian 
	distribution with mean $\mu$ and covariance $\Sigma$, and the uniform 
	distribution between $a$ and $b$, respectively; $\Xi_N(x)$ 
	denotes the Dirichlet sinc function $\Xi_N(x) =  
	\frac{\sin(\frac{N}{2}\pi x )}
	{N\sin(\frac{1}{2}\pi x )}$.
	
	\section{Near-Field Beam Split}\label{sec:2}
	In this section, we elaborate on the near-field beam-split effect in  
wideband ELAA systems. We will first introduce the system model based on the 
conventional hybrid beamforming architecture,  and then the near-field beam 
split effect is discussed from this model. 

\subsection{System Model}
				\begin{figure}
	\centering
	\includegraphics[width=3.5in]{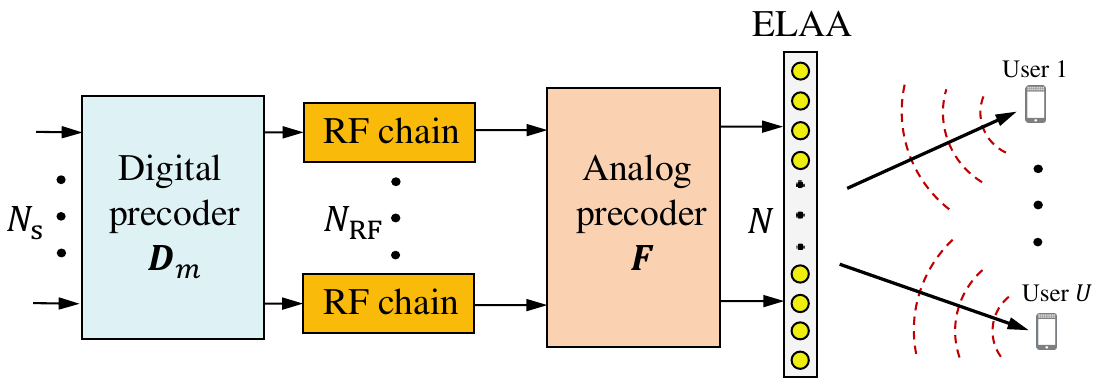}
	\caption{The system layout of ELAA.
	} 
	\label{img:layout}
	\vspace*{-1em}
\end{figure}

	In our system, a base station (BS) equipped with an $N$-element uniform 
	linear array (ULA)
 serves $U$ single-antenna users.  The BS employs the 
	fully 
	connected hybrid beamforming architecture with $N_{\rm RF}$ radio frequency 
	(RF) 
	chains, each connected to all antennas via analog
	phase shifters. To reap the multiplexing gain, $U$ data streams are 
	simultaneously transmitted, where $U \le N_{\rm RF}\ll N$. In this 
	article, we assume $U = N_{\rm RF}$ for illustration simplicity. 
	The orthogonal frequency division multiplexing (OFDM) with $M$ sub-carriers 
	is adopted. 
 Variables $d=\frac{\lambda_c}{2}$, $\lambda_c$, $c$, $f_c = 
 \frac{c}{\lambda_c}$, and $B$ denote the antenna spacing, carrier wavelength, 
 speed of light, center carrier, bandwidth, respectively. 	Besides, let $f_m = 
 f_c + \frac{B}{2}(\frac{2}{M-1}m - 1)$ for $m\in\{0,1,\cdots, M-1\}$ be the 
 $m^{\rm th}$ subcarrier frequency. {Accordingly, the  user-side 
 received 
 signal  $\mb{y}_m\in \mathbb{C}^{U\times 1}$ on the $m^{\rm th}$ subcarrier
 is expressed as 
 \begin{align}
 	\mb{y}_m = \sqrt{\rho} \mb{H}_m^T\mb{F}\mb{D}_{m} + 
 	\mb{n}_m, 
 \end{align}
where $\mb{H}\in\mathbb{C}^{N\times U}$, $\mb{F}\in\mathbb{C}^{N\times 
N_{\rm RF}}$, and $\mb{D}_{m}\in\mathbb{C}^{N_{\rm RF}\times U}$ 
represent 
the channel 
matrix, the analog beamformer, and the digital precoder respectively. It is 
noticeable that since the digital precoding is carried out 
subcarrier-by-subcarrier on the digital baseband, the digital precoder 
$\mb{D}_{m}$ 
is frequency-dependent.  However, the analog precoder realized by 
PSs can only tune a uniform phase for all subcarriers, making $\mb{F}$ 
frequency-independent. Taking into account the circuit restriction, each entry 
of  $\mb{F}$ satisfies the constant modulus constraint, i.e., 
$|[\mb{F}]_{i,j}| = \frac{1}{\sqrt{N}}$. The noise $\mb{n}_m$ follows 
the complex Gaussian distribution $\mathcal{CN}(0, \sigma^2\mb{I})$.}
 
We denote the channel matrix as $\mb{H}_m = [\mb{h}_{0,m}, \mb{h}_{1,m}, 
\cdots, \mb{h}_{U-1,m}]$, where vector $\mb{h}_{u,m}$ stands for the wireless 
channel from the BS to the $u^{\rm th}$ user.   
  As shown in Fig. 
 \ref{img:layout}, the center of the BS array is located at $(0,0)$ in 
 the Cartesian coordinates, and then the coordinate of the $n^{\rm th}$ BS 
 antenna is 
 $\left(0, 
 \delta_N^{(n)}d \right)$,
 where $\delta_N^{(n)} = n - \frac{N - 1}{2}$ with $n \in \{0, 1, \cdots, N - 
 1\}$.
 Therefore, the array aperture is given as $D = (N - 1)d\approx Nd$.
	The $u^{th}$ user is located at $(x_u, y_u)$, where its polar coordinate is 
	$(r_u, \theta_u) = \left(\sqrt{x_u^2 + y_u^2}, 
	\arctan{\frac{y_u}{x_u}}\right)$.  
	Then, adopting the free space Maxwell equation, the line-of-sight 
	near-field channel $\mb{h}_{u,m}$ 
	can be 
	modeled \cite{NearLoS_Zhou2015} as
	\begin{align}\label{eq:nf}
\mb{h}_{u,m}&= g_{u,m} \left[e^{-jk_mr^{(0)}_u}, e^{-jk_mr^{(1)}_u}, \cdots, 
e^{-jk_mr^{(N - 1)}_u} \right]^T \notag\\ 
&= g_{u,m}\sqrt{N} \mb{a}_m(r_u,\theta_u),
\end{align}
	where $\mb{a}_m(r_u,\theta_u)$ represents the near-field array response 
	vector, $k_m = \frac{2\pi f_m}{c}$ 
	denotes the wavenumber at frequency $f_m$, and $g_m$ denotes the complex
	path loss. Let 
	$r^{(n)}_u$ be the distance from the $n^{\rm th}$ BS antenna to the $u^{\rm 
	th}$ 
	user  
	expressed as 
	\begin{align}
r^{(n)}_u &= (x^2_u + (y_u - \delta_N^{(n)}d)^2)^{\frac{1}{2}} \notag\\&= 
(r_u^2 
+  
(\delta_N^{(n)}d)^2 - 2\delta_N^{(n)}dr_u\sin\theta_u)^\frac{1}{2}.
	\end{align}

				\begin{figure*}
			\centering
			\includegraphics[width=5in]{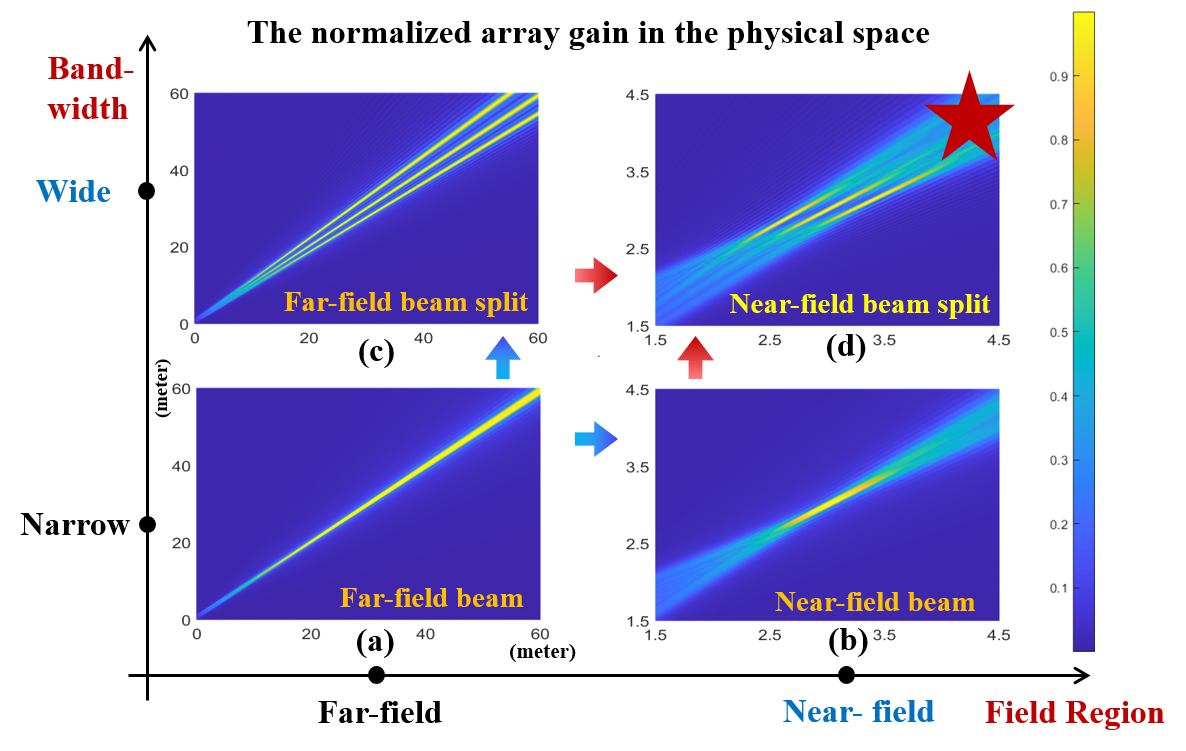}
			\caption{This figure illustrates the normalized beamforming gain in 
			the physical space.
				We consider four scenarios:
				(a) the far-field narrowband scenario, (b) the near-field 
				narrowband scenario, (c) the far-field wideband scenario, and 
				(d) the near-field wideband  scenario. In each sub-figure, the 
				beam energy of the lowest, the center, and the highest 
				frequencies are plotted (e.g., the three lines in the 
				sub-figures {(c)} and {(d)}).
			} 
			\label{img:FourBeamPattern}
			\vspace*{-1em}
		\end{figure*}
	
	Since the PS-based
	analog beamformer $\mb{F}$ is frequency-independent, each column of 
	$\mb{F}$ is generally set to align with the array response vector  of the 
	wireless channel at the center 
	frequency $f_c$ \cite{THzbeam_Headland2018, Overview_Heath2016}. 
 Thereafter, let $\mb{F} = 
	[\mb{f}_0, \mb{f}_1, \cdots, \mb{f}_{U-1}]$, then $\mb{f}_u$ can be 
	obtained 
	by  
	\begin{align}\label{eq:nw}
\mb{f}_u =	\mb{a}_c^*(r_u, \theta_u) = \frac{1}{\sqrt{N}}[
	 e^{jk_c r^{(0)}_u}, \cdots, e^{jk_c r^{(N-1)}_u}]^T,
		\end{align}
		where $k_c = \frac{2\pi f_c}{c}$ and $\mb{a}_c^*(r_u, \theta_u)$ 
		represent the wavenumber and array response vector on the 
		center frequency $f_c$.
	The fundamental model difference from the near-field array response vector 
	$\mb{a}_m(r_u, 
	\theta_u)$ to the analog beamformer $\mb{f}_u = \mb{a}_c^*(r_u, \theta_u)$ 
	causes the near-field beam split effect. 
	
	\subsection{Discussion on Near-Field Beam Split}

    To delve into the near-field beam split effect, we would 
	like 
	to 
	compare the beamforming properties under far-field/near-field 
	and 
	narrowband/widband conditions.  In the upcoming discussions, we will omit 
	the subscript $u$ for ease of
	expression. 
	Notice that the phase $k_m r^{(n)}$ in (\ref{eq:nf}) and $k_c r^{(n)}$ 
	(\ref{eq:nw}) are non-linear functions with respect to (w.r.t) the antenna 
	index $n$.  
	Traditionally, since the array aperture is not very large, the far-field 
	model under the planar wave assumption \cite{FundWC_Tse2015} is widely 
	adopted to simplify this non-linear distance as
	\begin{align}\label{eq:FF2NF}
	r^{(n)}  &= r\left(1 + \frac{(\delta_N^{(n)}d)^2}{r^2} - 
	\frac{2\delta_N^{(n)}d\sin\theta}{r}\right)^\frac{1}{2} \notag\\
	&\overset{ (a)}{\approx}   r\left(1 - 
	\frac{\delta_N^{(n)}d\sin\theta}{r}\right) = r -\delta_N^{(n)}d\sin\theta,
	\end{align}
	where (a) arises because of the first-order Taylor expansion $(1 + 
	x)^{\frac{1}{2}} \approx 1 + \frac{1}{2}x$ and the ignorance of the
	second-order term
	$\frac{(\delta_N^{(n)}d)^2}{r^2}$. 
	It is clear from (\ref{eq:FF2NF}) that in the far-field region, the phase 
	becomes $k_m r^{(n)} \approx  k_m r  - k_m\delta_N^{(n)}d\sin\theta$, which 
	is a linear function of the antenna index $n$. Then the far-field 
	beamforming vector becomes $[\mb{f}]_n = 
	\frac{1}{\sqrt{N}}e^{jk_cr}e^{-jk_c\delta_N^{(n)}d\sin\theta}.$ Because the 
	term $e^{jk_cr}$ is independent of the antenna index $n$, 
	$[\mb{f}]_n$ can be rewritten as  $[\mb{f}]_n = 
	\frac{1}{\sqrt{N}}e^{-jk_c\delta_N^{(n)}d\sin\theta} = 
	[\mb{f}^{\text{far}}(\theta)]_n$, depending only on direction $\theta$. 
	As shown in Fig. \ref{img:FourBeamPattern}\:(a), the beam at the center 
	frequency generated by $[\mb{f}^{\text{far}}(\theta)]_n$ is transmitting 
	towards a specific direction $\theta$.  
	
	However, since the linear approximation in (\ref{eq:FF2NF}) is not 
	accurate 
	when $n$ is very large, the above far-field assumption does not hold 
	anymore for ELAAs. The typical near-field range is 
	determined by the Rayleigh distance \cite{fresnel_Sherman1962} 
	$
	 R = \frac{2D^2}{\lambda_c} = \frac{1}{2}N^2\lambda_c.
	$ If the number of antennas $N$ increases dramatically, the near-field 
	region will expand by orders of magnitude. For instance, for a 512-antenna 
	ULA operating at 100 GHz frequency, the Rayleigh distance is about 400 
	meters. In this case, the accurate spherical wave model has to be adopted 
	for the channel $\mb{h}_m$ in (\ref{eq:nf}) and the 
	beamforming vector $\mb{f}$ in (\ref{eq:nw}). Accordingly, as 
	shown in Fig. \ref{img:FourBeamPattern}\:(b), the energy of near-field beam 
	at the center frequency is focused on the location $(r, \theta)$ 
	\cite{THzbeam_Headland2018}, which depends on both the distance $r$ and 
	direction 
	$\theta$ of 
	the receiver. Therefore, near-field beamforming also has the terminology 
	 ``beamfocusing" in the literature
	\cite{NearMag_Zhang23, NearBF_Zhang23}.


	The discussion above assumes that the bandwidth is not very large. As for 
	wideband systems, a severe beam split effect is induced. Specifically, 
	employing the beamforming vector $\mb{f} = \mb{a}_c^*(r, \theta)$, we 
	define 
	$G(\hat{r}, 
	\hat{\theta}, r, \theta, f_m) = |\mb{a}_m^T(\hat{r}, 
	\hat{\theta})\mb{f}|$  as the normalized beamforming 
	gain at 
	frequency 
	$f_m$ on the location $(\hat{r}, \hat{\theta})$ with $\hat{r}^{(n)} = 
	\sqrt{\hat{r}^2 + (\delta_N^{(n)}d)^2 - 
	2\delta_N^{(n)}d\hat{r}\hat{\theta}}$.  According to (\ref{eq:nf}), we 
	have  
	\begin{align} \label{eq:NF_gain}
	G(\hat{r}, \hat{\theta}, r, \theta, f_m) = \frac{1}{N}\left|\sum_{n=0}^{N-1} e^{-j (k_m\hat{r}^{(n)} -k_c r^{(n)}) }\right|.
	\end{align}
	Clearly, the maximum value of $G(\hat{r}, \hat{\theta}, r, \theta, f_m)$ is 
	1.
	In narrowband systems where $f_m \approx f_c$, the beamforming gain 
	$G(\hat{r}, \hat{\theta}, r, \theta, f_m)$ reaches its zenith when 
	$(\hat{r}, 
	\hat{\theta}) = (r, \theta)$, signifying that the beam energy is precisely 
	focused at location $(r,\theta)$. 
	However, in wideband systems, when $f_m \neq f_c$ and $(\hat{r}, 
	\hat{\theta}) = (r, \theta)$, the elements $ e^{-j(k_c - k_m)r^{(n)} 
	} $ exhibit diverse phases, preventing them from being contructively added 
	up on the 
	user location. This phenomenon substantially reduces the beamforming gain 
	$G(\hat{r}, \hat{\theta}, r, \theta, 
	f_m)|_{(\hat{r}, \hat{\theta})=(r,\theta)}$, rendering 
	it  much lower than 1. Consequently, the 
	beam energy at $f_m$ is split from the desired location $(r,\theta)$ 	
	\cite{NearBF_Cui23}. 
	{ In the far-field scenario where the distance is considerable,  
	as shown in 
	Fig. \ref{img:FourBeamPattern}\:(c), this beam split effect causes beams at 
	different frequencies to transmit towards \emph{different directions}.  
	However, in the near-field setting, as shown in Fig. 
	\ref{img:FourBeamPattern}\:(d), the near-field beam split effect causes 
	beams at 
	different frequencies to focus on \emph{different locations}. These 
	distinct beamforming properties distinguish the far-field and near-field 
	beam split effects\footnote{ Notice that the 
		sub-figures (c) and (d) of Fig. \ref{img:FourBeamPattern} also appear 
		in our earlier work~\cite{NearBF_Cui23}, where \cite{NearBF_Cui23} 
		employs  near-field 
		beam split to perform fast beam training, while the objective of this 
		work is to mitigate this phenomenon via energy-efficient beamforming. 
}.
} 

Furthermore, since beams over large bandwidth are split to different 
locations/directions, the user can only access to signals close to the center 
frequency.
	For example, considering $f_c = 100$ GHz, $B = 5$ GHz, and $N = 512$, 
	the near-field beam-split effect results in over $50\%$ of the 
	sub-carriers experiencing a beamforming gain loss of at least $60\%$.
	Recent works mainly concentrate on mitigating the far-field beam-split 
	effect. This is accomplished by either deploying a large number of high 
	power consumption time-delay elements \cite{TTDhardware_Hashemi2008}, 
	\cite{TTD_Lin2017} or heavily  relying on the linear phase property of the 
	far-field channel 
	\cite{DPP_Tan2019, BSprecoding_Hanzo2020, STBC_Liu2019,MOALT_Yu2016}, which 
	is not applicable in the near field.
	To the best of our knowledge, the near-field beam-split effect has not been 
	studied in the literature. 
	\section{Proposed Methods}\label{sec:3}
	In this section, we commence by introducing a piecewise-far-field channel 
	characterized by piecewise-linear phase properties to approximate the 
	intricate near-field 
	channel. Subsequently, a PDF method is built on this approximation to 
	mitigate the near-field beam split effect. 
		\begin{figure*}
		\centering
		\subfigure[Far-field channel]{
			\begin{minipage}[t]{0.15\linewidth}
				\centering
				\includegraphics[width=1in,height=1.7in]{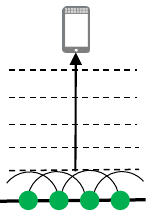}
			\end{minipage}%
		}%
		\subfigure[Near-field channel]{
			\begin{minipage}[t]{0.3\linewidth}
				\centering
				\includegraphics[width=2in,height=1.8in]{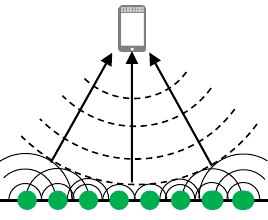}
			\end{minipage}%
		}%
		\subfigure[Proposed piecewise-far-field channel]{
			\begin{minipage}[t]{0.3\linewidth}
				\centering
				\includegraphics[width=2in,height=1.8in]{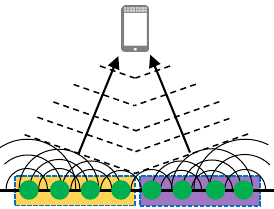}
			\end{minipage}
		}%
		\\
		\centering
		\subfigure[layout of the piecewise-far-field model
		]{\includegraphics[width=3in]{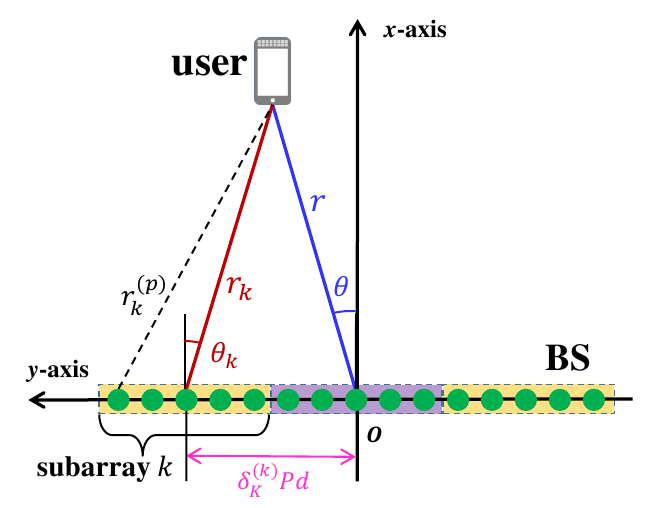}}
		\subfigure[Channel phase against antenna index 
		]{\includegraphics[width=3in]{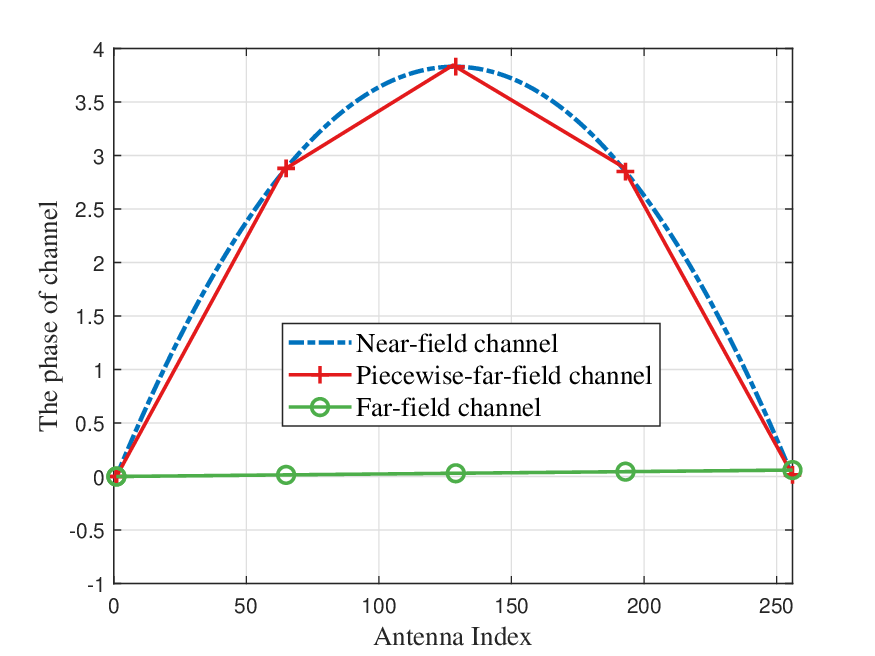}}
		\caption{ Schematic diagrams of {(a)} far-field channel model, {(b)} 
		near-field channel model, {(c)} piecewise-far-field channel model,
			{(d)} layout of the piecewise-far-field model, and {(e)} channel 
			phase against the antenna index. The number of 
			antennas 
			is 256, the carrier frequency is 100 GHz. The user is located at 
			$(x,y) 
			= (10 \: \text{m}, 0\: \text{m})$. With $K = 4$ sub-arrays, the 
			piecewise-far-field channel model can well approximate the 
			near-field 
			channel model.
		}
		\label{img:group}
		\vspace*{-1em}
	\end{figure*}
	\subsection{Piecewise-Far-Field Channel Model}\label{sec:3.1}
    The non-linear phase $-k_mr^{(n)}_u$ w.r.t the antenna index $n$ makes it 
    intractable to 
    directly devise near-field wideband beamforming techniques. In order to get 
    a manageable simplification of this non-linear phase while maintaining  
    acceptable accuracy, we observe that the 
    Rayleigh distance $ \frac{1}{2}N^2\lambda_c $ scales proportionally with 
    the square 
    of the number of antennas, signifying that fewer antennas corresponds to a 
    better 
    accuracy of the far-field assumption in (\ref{eq:FF2NF}). 
    
	Inspired by this observation, as depicted in Fig. \ref{img:group}\:(a)-(c), 
	a 
	piecewise-far-field channel model is harnessed to approximate 
	the intricate near-field channel. In this model, the entire large array is 
	partitioned into multiple sub-arrays, each equipped with much fewer 
	antennas compared to the entire array. This partition leads to a notable 
	reduction of the near-field range for each sub-array.
	Consequently, even if the receiver is inside the near-field 
	region of 
	the entire array, we can reasonably assume that the receiver is 
	situated in 
	the far-field 
	region of each sub-array. 
	To elaborate, as shown in Fig. \ref{img:group}\:(d), we divide the entire 
	large array into $K$ sub-arrays. For each sub-array, there are $P$ adjacent 
	antennas, satisfying $N = KP$. Then, 
	the near-field channel from the BS to the $u^{\rm th}$ user is 
	rearranged as follows
	\begin{align} \label{eq:decom}
	\mb{h}_{u, m} = \left[ {\mb{h}_{u,m}^{\left(0\right)^T}}, 
	{\mb{h}_{u,m}^{\left(1\right)^T}},\cdots,{\mb{h}_{u,m}^{\left(K-1\right)^T}}
	 \right]^T, 
	\end{align}
	where $\mb{h}_{u,m}^{(k)} \in \mathbb{C}^{P \times 1}$ represents the 
	sub-channel 
	between the $k^{\rm th}$ sub-array and the $u^{\rm th}$ user.  We define  
	$\delta_K^{(k)} = k 
	- \frac{K - 1}{2}$ for $k \in \{0,1,\cdots, K -1\}$. Then, the distance and 
	direction from the center of the $k^{\rm th}$ sub-array to the $u^{\rm th}$ 
	user 
	are expressed as $r_{u,k} = 
	\sqrt{x_u^2 + (y_u - \delta_K^{(k)}Pd)^2} = \sqrt{r_u^2 + 
	(\delta_K^{(k)}Pd)^2 - 
	2\delta_K^{(k)}Pdr_u\sin\theta_u}$ and $\sin\theta_{u,k} = \frac{y_u - 
	\delta_K^{(k)}Pd}{r_{u,k}}$, respectively. We further define 
	$\delta_P^{(p)} = p - 
	\frac{P 
	- 1}{2}$ for $p \in \{0,1,\cdots, P -1\}$. Subsequently, according to 
	(\ref{eq:FF2NF}),  
	the distance $r_{u,k}^{(p)}$ from the $p^{\rm th}$ antenna in the $k^{\rm 
	th}$ 
	sub-array 
	to the $u^{\rm th}$ user is expressed as 
	 \begin{align}
	 r_{u,k}^{(p)} & = \sqrt{r_{u,k}^2 + (\delta_P^{(p)}d)^2 - 
	 2\delta_P^{(p)}dr_{u,k}\sin\theta_{u,k} } \notag\\
	 & \overset{(a)}{\approx} r_{u,k} - \delta_P^{(p)}d\sin\theta_{u,k}.
	 \end{align}
 There, the approximation (a) holds because each sub-array is 
 small enough.
Accordingly, the near-field channel  $\mb{h}_{u,m}^{(k)}$ of the $k^{\rm th}$ 
sub-array
is approximated by a far-field channel $\tilde{\mb{h}}_{u,m}^{(k)}$:
	\begin{align} \label{eq:PiecewiseLinear}
[\mb{h}_{u,m}^{(k)}]_p & \approx g_{u,m} e^{-jk_m r_{u,k}} e^{jk_m 
\delta_P^{(p)}d\sin\theta_{u,k}}  = [\tilde{\mb{h}}_{u,m}^{(k)}]_p.
\end{align} 
By introducing the parameter $\eta_m = \frac{f_m}{f_c}$ and plugging $d = 
\frac{\lambda_c}{2} = 
\frac{c}{2f_c}$ and $k_m = \frac{2 \pi f_m}{c}$ into 
(\ref{eq:PiecewiseLinear}), we arrive at  
$[\tilde{\mb{h}}_{u,m}^{(k)}]_p = e^{-jk_m r_{u,k}} e^{j\pi \eta_m 
\delta_P^{(p)}\sin\theta_{u,k}}$.
Consequently, the intricate near-field channel is approximated by a 
piecewise-far-field channel:
	\begin{align} \label{eq:PFFchannel}
	\mb{h}_{u,m} \approx  \tilde{\mb{h}}_{u,m} = \left[ 
	\tilde{\mb{h}}_{u,m}^{\left(0\right)}, 
	\tilde{\mb{h}}_{u,m}^{\left(1\right)},\cdots,\tilde{\mb{h}}_{u,m}^{\left(K 
	- 
	1\right)} \right].
	\end{align}
It is notable from \eqref{eq:PiecewiseLinear} that the phase of 
$[\tilde{\mb{h}}_{u,m}^{(k)}]_p$ is a linear 
function of $p$, the antenna index of the $k^{\rm th}$ sub-array. This linear 
phase property suggests that $[\tilde{\mb{h}}_{u,m}^{(k)}]_p$ can be regarded 
as 
a far-field channel. 
	Furthermore, considering the different $r_k$ and $\theta_k$ in each 
	sub-array, the planar waves impinging on different 
	sub-arrays come from different directions. This is why we call the entire 
	array's channel the piecewise-far-field channel. 
	To illustrate the fidelity of the proposed model,  Fig. 
	\ref{img:group}\:(e) 
	depicts the channel phase as a function of the antenna index for the 
	near-field, far-field, and piecewise-far-field channel models. 
	The phase profile of the piecewise-far-field channel model closely 
	approaches  that of the true near-field channel. 
	In essence, our proposed channel 
	model can be recognized as a piecewise-linearization of the intricate 
	near-field 
	channel 
	model, wherein the phase exhibits local linear behavior within each 
	sub-array.
	Harnessing this piecewise-linear phase characteristics, we proceed to 
	devise a near-field wideband beamforming method referred 
	to as phase-delay focusing to alleviate the near-field beam-split effect in 
	the subsequent subsection.
	\subsection{Proposed Phase-Delay Focusing Method}\label{sec:3.2}  

	We first elaborate on overcoming the near-field beam-split 
	for an arbitrary user by analog beamforming, while the extension to 
	multi-user hybrid beamforming is studied in Section \ref{sec:3.4}. For 
	ease of expression, the subscript $u$ is omitted in Section \ref{sec:3.2} 
	to \ref{sec:3.3}. Accordingly, the variables $\mb{h}_{u,m}$, 
	$\tilde{\mb{h}}_{u,m}$,
	$\mb{f}_u$, $r_u$, $\theta_u$, $r_{u,k}$, and $\theta_{u,k}$ become  
$\mb{h}_{m}$, 
$\tilde{\mb{h}}_{m}$,
$\mb{f}$, $r$, $\theta$, $r_{k}$, and $\theta_{k}$.
	
   	Specifically, the introduced piecewise-far-field channel model makes it 
   	straightforward  
   	to 
   	decouple the phase in (\ref{eq:PiecewiseLinear}) into two components: the 
   	inter-array phase discrepancy $ - k_m r_{k}$ across different 
   	sub-arrays, and the intra-array phase discrepancy $ \pi 
   	\eta_m \delta_P^{(p)} \sin \theta_{k} $ within each sub-array. It is 
   	notable 
   	that $ - k_m r_{k}$ is a non-linear function of $k$, giving rise to a 
   	near-field channel phase property, whereas $ \pi 
   	\eta_m \delta_P^{(p)} \sin \theta_{k} $ follows a linear function in 
   	relation 
   	to $p$, as the 
   	same to a far-field model. 
   	Both of these two phase components contribute to the near-field beam split 
   	effect.  The following fact inspires us to neglect the 
   	influence of 
   	intra-array phase $\pi \eta_m \delta_P^{(p)} \sin \theta_{k}$
   	on near-field beam split. As suggested in \cite{SW_Wang2018, 
   	DPP_Tan2019}, 
   	the degree of beam-split effect 
   	is 
   	proportional to the physical antenna aperture. A larger 
   	antenna aperture results in a severer beam-split effect. Although this 
   	conclusion is derived in the far-field region, it is valid to 
   	near-field as well, because the physical propagation delay always increases 
   	with the antenna aperture, no matter far-field or near-field. Following 
   	this intuition, we can find that the 
   	intra-array phase discrepancy $k_m 
   	\delta_P^{(p)}d\sin\theta_{k}$ corresponds to a sub-array's aperture 
   	${Pd}$, while the inter-array phase 
   	discrepancy $k_m r_k$ is related 
   	to the entire array's aperture $Nd$. Since 
   	$P\ll N$, it is reasonable to deduce the near-field beam split is dominated 
   	by the inter-array phase discrepancy. 
   	 Consequently, our target is converted to compensating  for 
   	the inter-array phase $ - k_m r_{k}$. 
   	
   			\begin{figure}
   		\centering
   		\includegraphics[width=3.5in]{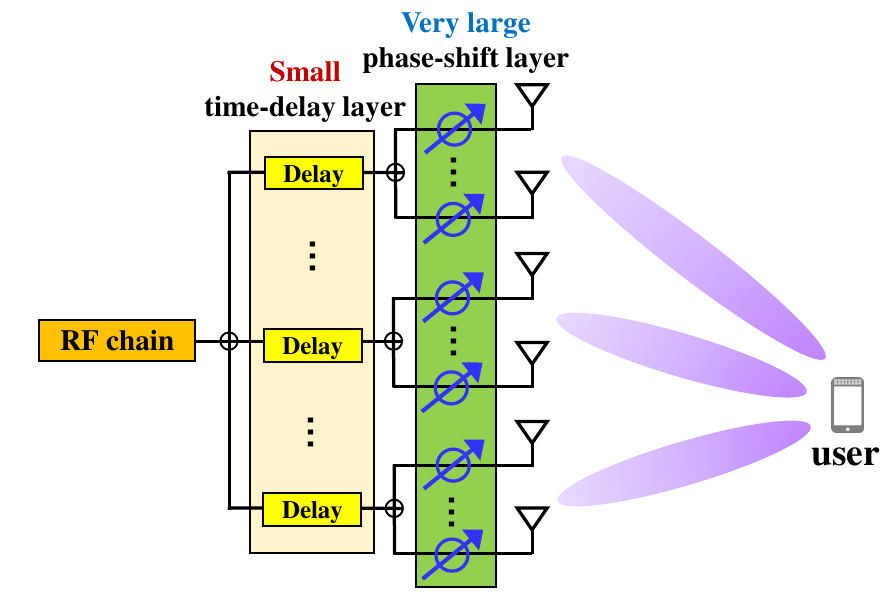}
   		\caption{A single-user example of the delay-phase precoding 
   		architecture \cite{DPP_Tan2019} for performing the 
   		PDF method.
   		}
   		\label{img:DPP}
   		\vspace*{-1em}
   	\end{figure}
   
    Note that the channel phase $ - k_m r_{k} = - \frac{2\pi f_m r_{k}}{c}$ 
    is equivalent to the frequency response of a time 
    delay of 
    $\frac{r_k}{c}$. Therefore, the delay-phase precoding architecture  
    \cite{DPP_Tan2019} employing TTD circuits can be used to compensate for the 
    inter-array phase $ - k_m r_k$. As illustrated in Fig. 
    \ref{img:DPP}, compared to conventional hybrid precoding architecture,  one 
    additional TTD circuit is inserted in each sub-array to connect the RF 
    chain 
    and the 
    PS-based sub-array. The frequency 
    response of a TTD at $f_m$ is $e^{-j2\pi f_m \tau'}$, where $\tau'$ 
    represents the 
    adjustable time delay parameter. For brevity, let $r' = 
    c\tau'$ be the adjustable distance parameter. Then, the 
    corresponding 
    frequency response is transformed into $e^{-jk_mr'}$. Thereby, a TTD is 
    able 
    to compensate for the frequency-dependent phase $- k_m r_k$ if $r' = -r_k$. 
    Moreover, as shown in Fig. \ref{img:DPP}, the main function of the PS-based 
    sub-array is to generate far-field planar waves to match the  intra-array 
    far-field 
    phase  $ \pi \eta_m \delta_P^{(p)}\sin \theta_k$. 
    Through the joint manipulation of PSs and TTDs, the beam energy across the 
    entire bandwidth can be focused on the receiver location 
    $(r,\theta)$. We henceforth refer to this method as phase-delay focusing.

	 Recall that the analog beamformer $\mb{f}$ realized by PS is 
	 frequency-independent. In contrast, the introduction of TTD makes the 
	 corresponding analog beamformer $\mb{f}_m$ frequency-dependent.  
	 To be specific, similar to the decomposition of wireless channel presented 
	 in (\ref{eq:decom}), 
	 the beamforming vector $\mb{f}_m$ realized by 
	 the PDF method at $f_m$ is composed of $K$ sub-vectors, i.e., 
	\begin{align}\label{eq:ww}
	\mb{f}_m = \left[
	{\mb{f}_m^{\left(0 \right)^T}} ,
	{\mb{f}_m^{\left(1 \right)^T}} ,\cdots,
	{\mb{f}_m^{\left(K - 1 \right)^T}}  \right]^T,
	\end{align}
	where $\mb{f}_m^{(k)} \in \mathbb{C}^{P\times 1}$ represents the 
	beamforming 
	vector of the $k^{\rm th}$ sub-array. As illustrated in Fig. \ref{img:DPP}, 
	$\mb{f}_m^{(k)}$ is generated by one TTD element and $P$ PSs, which can be 
	expressed as follows:
	\begin{align}\label{eq:wwk}
	\mb{f}_m^{(k)} = \frac{1}{\sqrt{N}} e^{-jk_m r_k' } [e^{ j\pi 
	\delta_P^{(0)} \beta_k' }, 
	\cdots, e^{ j\pi \delta_P^{(P-1)} \beta_k' }]^T.
	\end{align} 
	Here, $r_k'$ denotes the adjustable distance parameter of the $k^{\rm th}$ 
	TTD 
	element and  $\beta_k'$ denotes the adjustable phase parameter associated 
	with 
	the $k^{\rm th}$ PS-based sub-array.
	It is evident from (\ref{eq:wwk}) that $\mb{f}_m^{(k)}$ includes two 
	distinct 
	components. The first one is the frequency-independent phase $\pi 
	\delta_P^{(p)}\beta_k'$ generated by the $P$ 
	PSs within the $k^{\rm th}$ sub-array. The second one involves the 
	frequency-dependent phase $-k_m r_k'$ realized by the $k$th TTD element.
    As discussed earlier, the purpose of $\pi \delta_P^{(p)}\beta_k'$ is  to 
    produce 
    planar waves that align with the far-field phase discrepancy $ \pi \eta_m 
    \delta_P^{(p)}\sin 
    \theta_k$,
    while the introduction of $t_k' = \frac{r_k'}{c}$ serves to compensate for 
    the near-field phase discrepancy $ - k_m r_k$. 
    
 To elaborate, the normalized beamforming gain achieved by the proposed PDF 
 method on the user location is expressed as
    \begin{align} \label{eq:arg2}
    G_m &= 
    \frac{1}{\sqrt{|g_m|N}}|\tilde{\mb{h}}_m^T\mb{f}_m| =\frac{1}{\sqrt{|g_m|N}}
    \left|\sum_{k=0}^{K-1}\tilde{\mb{h}}_m^{(k)^T}\mb{f}_m^{(k)}\right| \notag 
    \\
    &	= \frac{1}{N} \left|\sum_{k=0}^{K - 1} e^{-jk_m (r_k' + r_k)} \sum_{p = 0 }^{P - 1} e^{j\delta_P^{(p)}\pi(\beta_k' + \eta_m \sin\theta_k)} \right|
    \notag\\ 
    & = \frac{1}{K}\left|\sum_{k=0}^{K - 1} e^{-jk_m (r_k' + r_k)} 
    \Xi_P(\beta_k' + \eta_m \sin\theta_k)\right|,
    \end{align}
    where $\Xi_P( x ) = \frac{\sin\left(\frac{P}{2}\pi x\right)} {P 
    \sin\left(\frac{1}{2}\pi x\right)}$.
    To generate planar waves aligning with the sub-array channel, $\beta_k'$ is 
    typically devised according to the spatial direction $\sin\theta_k$ at the 
    center frequency \cite{PhaseArray_Mailloux2005}, i.e., 
    \begin{align} \label{eq:beta_k}
	\beta_k' = - \sin\theta_k. 
\end{align}
    By substituting (\ref{eq:beta_k}) into (\ref{eq:arg2}), we 
    obtain 
    \begin{align} \label{eq:arg}
    &\tilde{\mb{h}}_m^{(k)^T}\mb{f}_m^{(k)} = 	 e^{-k_m (r_k' + r_k)}
    \Xi_P( \epsilon_m\sin\theta_k),
    \end{align}
where $\epsilon_m = \eta_m - 1 = \frac{B}{f_c}(\frac{2}{M-1}m - 1)$. 
    The subsequent objective of our PDF method is to find proper 
    $\{r_k'\}$ to 
    maximize the beamforming 
    gain on the user location $(r,\theta)$ over the entire bandwidth. Hence,  
    the corresponding optimization problem can be 
    formulated as 
    \begin{align} \label{eq:ta}
    \mathop{\mbox{max}}\limits_{\{r_k\}} \quad &\frac{1}{MK} 
    \sum_{m=1}^M\left|\sum_{k=0}^{K - 1} e^{-jk_m (r_k' + r_k)} 
    \Xi_P(\epsilon_m \sin\theta_k) \right| \\
    \mbox{s.t.}\quad
    & r_k' \ge 0 \quad  k \in \{1,2,\cdots,K\}.  \notag 
    \end{align}
    We provide the following \textbf{Lemma \ref{lemma:1}} to solve problem 
    (\ref{eq:ta}).
    \begin{lemma}\label{lemma:1}
    	If $|\epsilon_m| \le \frac{2}{P}$ for  $\forall m \in 
    	\{0,1,\cdots,M-1\}$, then the optimal solution to problem (\ref{eq:ta}) 
    	is
    	\begin{align}
	    	& r_k'
	    	= L - r_k , \label{eq:t}
	    	\end{align}
    	where $L$ is a global distance parameter chosen to ensure 
    	$\min\{r_k'\}\ge0$.
    \end{lemma}
    \begin{IEEEproof}
    	By substituting (\ref{eq:t}) into  (\ref{eq:arg}), the beamforming gain 
    	of the $k^{\rm th}$ sub-array can rewritten as 
    	\begin{align}
    	\tilde{\mb{h}}_m^{(k)^T}\mb{f}_m^{(k)} =  P
    	e^{-jk_m L} 	\Xi_P( \epsilon_m\sin\theta_k). 
    	\end{align}
    	The condition $ |\epsilon_m| \le 
    	\frac{2}{P}$ means that the parameters $|\epsilon_m\sin\theta_k| \le 
    	|\epsilon_m|$ 
    	are within the main lobe of $\Xi_P(\cdot)$ for all subcarriers $f_m$.
    	Hence, it is obvious that $\Xi_P(\epsilon_m\sin\theta_k)>0$, and    	
    	the beamforming gain can be presented as
    	\begin{align}\label{eq:an1}
    	\frac{1}{N}\left|\sum_k\tilde{\mb{h}}_m^{(k)^T}\mb{f}_m^{(k)} \right|
    	= \frac{1}{K}\sum_k	\Xi_P( \epsilon_m\sin\theta_k).
    	\end{align}
    	In addition, according to the Cauchy-Schwarz inequality, the 
    	beamforming gain $G_m$ has an upper bound: 
    	\begin{align} \label{eq:an2}
    	G_m \le \frac{1}{K}	\sum_k\left| \tilde{\mb{h}}_m^{(k)^T}\mb{f}_m^{(k)} 
    	\right|
    	= \frac{1}{K}\sum_k	\Xi_P( \epsilon_m\sin\theta_k).
    	\end{align}
    	It is clear from (\ref{eq:an1}) and (\ref{eq:an2}) that  $r_k'$ is the 
    	optimal solution to maximize $G_m$ at frequency $f_m$.
    	Moreover, since $r_k'$ is frequency independent, it is the optimal 
    	solution to all sub-carriers and thus optimal to problem (\ref{eq:ta}),
    	which completes the proof.
    \end{IEEEproof}

 To sum up, (\ref{eq:beta_k}) and (\ref{eq:t}) complete the beamforming design 
 for the proposed PDF method. In the next subsection, 
 theoretical analysis will be provided to validate the efficiency of the 
 proposed PDF method in 
 alleviating the near-field beam split effect.
 
 \subsection{Analysis of Beamforming Gain Performance}\label{sec:3.3}
This subsection presents an analysis of the performance of the 
proposed PDF method for large numbers $M$ and $K$.  We begin by introducing 
\textbf{Lemma \ref{lemma:2}}, which provides 
the average 
 beamforming gain over all sub-carriers 
 achieved by the PDF method.
 
  \begin{lemma} \label{lemma:2}
 	Suppose the user is located in the far-field region of each sub-array and 
 	$|\epsilon_m| \le \frac{2}{P}$ for $\forall m$, then the 
 	average beamforming gain over all sub-carriers achieved by (\ref{eq:ww}) 
 	can be 
 	approximated as
 	\begin{align}
 		\frac{1}{M}\sum_{m=0}^{M-1}G_m \approx   1 - \frac{1 - 
 		\Xi_P(\frac{B}{2f_c})}{MK} \sum_{m = 
 		0}^{M-1}
 		\frac{\epsilon_m^2}{(\frac{B}{2f_c})^2} \sum_{k = 0}^{K-1} 
 		\sin^2\theta_k.
 		\label{eq:L2}
 	\end{align}
 \end{lemma}
\begin{IEEEproof}
	 	\begin{figure}
		\centering
		\includegraphics[width=3.5in]{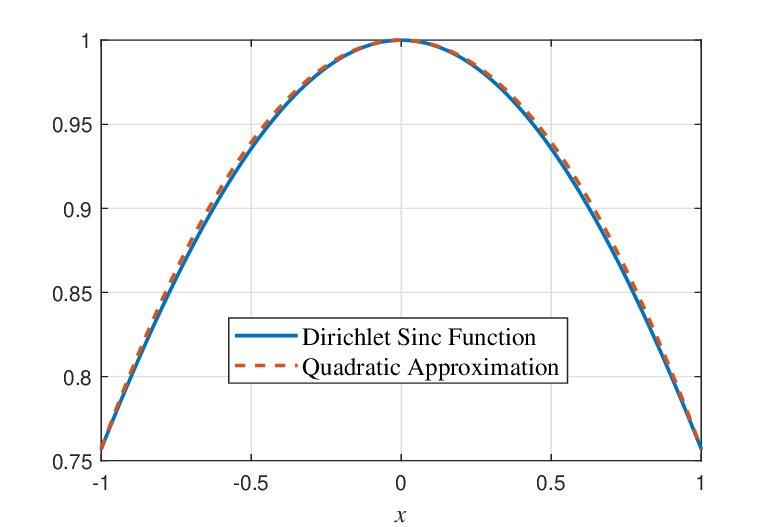}
		\caption{Quadratic fitting $1 - \left(1 -  
			\Xi_P(\frac{B}{2f_c})\right)x^2$ of the Dirichlet sinc function 
			$\Xi_P(\frac{B}{2f_c}x)$, where $x\in[-1,1]$, $P = 32$, $B = 5$ 
			GHz, 
			and $f_c = 100$ GHz.}
		\vspace*{-1em}
		\label{img:Quadratic approximation}
	\end{figure}
	Plugging (\ref{eq:beta_k}) and (\ref{eq:t}) into (\ref{eq:arg2}), the 
	average beamforming gain could be expressed as 
	 	\begin{align} \label{eq:SGm}
		\frac{1}{M}\sum_{m=0}^{M-1}G_m = \frac{1}{MK} 
		\sum_{m=0}^{M-1}\sum_{k = 0}^{K-1}  \Xi_P(\epsilon_m \sin\theta_k). 
	\end{align}
	It is intractable to compute (\ref{eq:SGm}) as the variables $\epsilon_m$ 
	and $\sin\theta_k$ are included in Dirichlet sinc functions 
	$\Xi_P(\epsilon_m \sin\theta_k)$. To deal with 
	this problem, we use a two-variable quadratic function to fit function
	$\Xi_P(ab)$. To be specific, because $|\epsilon_m| \le \epsilon_{M - 1} = 
	\frac{B}{2f_c}$ and 
	$|\sin\theta_k| \le 1$, the following five points on $\Xi_P(ab)$ are used 
	for function fitting: $(0,0,1)$, $(\frac{B}{2f_c}, 1, 
	\Xi_P(\frac{B}{2f_c}))$, $(\frac{B}{2f_c}, -1, \Xi_P(\frac{B}{2f_c}))$, 
	$(-\frac{B}{2f_c}, 
	1, \Xi_P(\frac{B}{2f_c}))$, $(-\frac{B}{2f_c}, -1, \Xi_P(\frac{B}{2f_c}))$. 
	Therefore, 
	we have 
	\begin{align} \label{eq:ApproSinc}
		\Xi_P(ab) \approx 1 - (1 - 
		\Xi_P(\frac{B}{2f_c}))   \frac{a^2}{(\frac{B}{2f_c})^2} b^2. 
	\end{align}
	The graph of function $1 - (1 - \Xi_P(\frac{B}{2f_c}))x^2$ is depicted in 
	Fig.~\ref{img:Quadratic approximation} for parameters $P = 32$, $B = 5$ 
	GHz, and $f_c = 100$ GHz, which is quite close to the Dirichlet sinc 
	function $\Xi_P(\frac{B}{2f_c}x)$. 
	Finally, substituting (\ref{eq:ApproSinc}) into (\ref{eq:SGm}) and 
	replacing 
	$(a,b)$ 
	with $(\epsilon_m, \sin\theta_k)$, we could arrive at the conclusion 
	(\ref{eq:L2}).
\end{IEEEproof}
\textbf{Lemma \ref{lemma:2}} allows us to separately compute the factors 
affecting the 
average beamforming gain:  the factor $\sum_m 
\frac{\epsilon_m^2}{(\frac{B}{2f_c})^2}$ 
arising from the wideband effect and the factor $\sum_k 
\sin^2\theta_k$ 
resulting from the near-field effect. 

Specifically, the following \textbf{Corollary 1} presents a more analytical 
form 
of (\ref{eq:L2}). 
  \begin{corollary} \label{Corollary:2}
  	For large numbers $K$ and $M$, the average beamforming gain $G = 
  	\frac{1}{M}\sum_{m=0}^{M-1}G_m $ in (\ref{eq:L2}) could be represented as
  	\begin{align}
  		G = \frac{1}{M}\sum_{m=0}^{M-1}G_m \approx 1 - \gamma(B, f_c, P) \times 
  		\xi(r, 
  		\theta, D), \label{eq:C2}
  		\end{align}
  	where 
  	\begin{align}
\gamma(B, f_c, P) &= 
  	\frac{1 
  		- 
  		\Xi_P(\frac{B}{2f_c})}{3}, \\
  	  		\xi(r, 
  	\theta, D) &= 
  	1 - \frac{r\cos\theta}{D} \left(\pi \mathbb{I}_{2r \le D} + \arctan 
  	\frac{Dr\cos\theta}{r^2 - \frac{1}{4}D^2}\right).
  	\end{align}
  Here,  $D = (KP - 1)d \approx KPd$, $\mathbb{I}_{2r \le D} = 1$ if $2r \le 
  D$, 
  and $\mathbb{I}_{2r \le D} = 0$ 
  otherwise.
\end{corollary}

\begin{IEEEproof}
The key to this proof lies in calculating the close-form expressions of $\sum_m 
\frac{\epsilon_m^2}{(\frac{B}{2f_c})^2}$ and $\sum \sin^2\theta_k$. 
Firstly, due to the fact that  $1^2 + 2^2 + \cdots + n^2 = 
\frac{n(n+1)(2n+1)}{6}$, we can represent  $\sum_m 
\frac{\epsilon_m^2}{(\frac{B}{2f_c})^2}$ as 
\begin{align}\label{eq:sm}
	\sum_m \frac{\epsilon_m^2}{(\frac{B}{2f_c})^2} = \sum_{m = 
		0}^{M-1} 
	\left(\frac{2}{M-1}m 
	- 1\right)^2 = \frac{M(M + 1)}{3(M - 1)},
\end{align}
Then for a large number $M$,   $\sum_m 
\frac{\epsilon_m^2}{(\frac{B}{2f_c})^2}$ could be further approximated as 
$\frac{M}{3}$. 

As for the second summation $\sum_k \sin^2\theta_k$, it could be rewritten in 
an 
integral form for a large number $K$: 
\begin{align}\label{eq:sk}
	&\sum_{k = 0}^{K-1} \sin^2\theta_k = \sum_{k = 0}^{K-1} 
	\frac{(r\sin\theta - 
		\delta_K^{(k)}Pd)^2}{r^2 + (\delta_K^{(k)}Pd)^2 - 
		2\delta_K^{(n)}Pdr\sin\theta} \notag\\
	&\approx\int_{-\frac{K}{2}}^{\frac{K}{2}}\frac{(r\sin\theta - 
		kPd)^2}{r^2 + (kPd)^2 - 
		2kPdr\sin\theta} \text{d}k \notag\\
	& \overset{(a)}{=} K - \frac{r\cos\theta}{Pd}\left.\arctan\frac{Pdk - 
		r\sin\theta}{r\cos\theta} \right|_{-K/2}^{K/2} \notag\\
	& \overset{(b)}{=}   	K - \frac{r\cos\theta}{Pd} \left(\pi \mathbb{I}_{2r 
	\le D} + \arctan 
	\frac{Dr\cos\theta}{r^2 - \frac{1}{4}D^2}\right).
\end{align}
where $D = (KP-1)d \approx KPd$ for large $KP$. Here, (a) comes from the 
indefinite integral 
$\int\frac{\text{d}x}{A+2Bx+Cx^2} = \frac{1}{\sqrt{AC-B^2}}\arctan \frac{Cx 
+B}{\sqrt{AC-B^2}}$ for $AC > B^2$ \cite{Table_2014}, and (b) arises from a 
characteristic of the 
inverse 
tangent function:
 \begin{equation*}\arctan A + \arctan B = \left\{
 	\begin{array}{lr}
 		\arctan\frac{A + B}{1 - AB} & (AB < 1) \\
 		\pi + \arctan\frac{A + B}{1 - AB} & (AB > 1) \\
 	\end{array} 
 \right.
 \end{equation*}
for $A > 0$.
Finally, by combining (\ref{eq:L2}), (\ref{eq:sm}), and (\ref{eq:sk}), 
the conclusion of \textbf{Corollary 1} could be  obtained. 
\end{IEEEproof}

\textbf{Corollary 1} yields several crucial conclusions. The beamforming gain 
loss of our PDF method arises from two factors: the loss caused by wideband 
effect $\gamma(B, 
f_c, P)$ and the loss posed by the geometry $\xi(r, \theta, D)$. 

First, the geometry loss $\xi(r, \theta, D)$ captures the degree of beam 
split effect with varying user locations. Due to the non-decreasing property 
of function $\arctan(.)$, it is straightfoward to prove that $\xi(r, \theta, 
D)$ monotonically increases w.r.t $|\theta|$ when $r > 
\frac{1}{2}D$. This fact implies 
that a larger angle of arrival leads to a more significant beamforming gain 
loss caused by near-field beam split, making it harder for our PDF 
method to compensate for this loss. Similar conclusions also appear in existing 
far-field beam split solutions. Moreover, $\xi(r, \theta, D)$ could account for 
the geometry loss in both far-field and near-field regions, as it incorporates 
the distance
parameter $r$. A simple evidence is that when 
$r\rightarrow+\infty$, $  	
\xi(r, \theta, D)$ tends to $\sin^2\theta$, which is exactly the geometry loss 
in far-field conditions provided in \cite{DPP_Tan2019}.

Second, $\gamma(B, f_c, P)$ is induced by the loss of beam split within each 
sub-array. This is due to the employment of PS based frequency-independent 
beamforming for each sub-array. Fortunately, since the number $P$ of a 
sub-array's 
antennas is much less than the number $N$ of the entire array's antennas, this 
loss 
$\gamma(B, f_c, P)$ can approach 0 by choosing an appropriate value of $P$. 
To be specific, it can be easily proven that $\gamma(B, f_c, P)$ is an 
increasing function w.r.t 
$P$ when $P\le \frac{4f_c}{B}$. With a smaller $P$, the impact of 
intra-array beam split is reduced. 
For instance, when $P = 
32$, $B = 5$ GHz, $f_c = 100$ GHz, $r = 10$ m, $D = 0.5$ m, and $\theta = 
\frac{\pi}{3}$, we have $\gamma(B, f_c, P) \approx 0.081$, $\xi(r, \theta, D) 
\approx 0.7496$, and $1 - \gamma(B, f_c, P) \times \xi(r, \theta, D) 
\approx 0.9393$. This implies that more than $93\%$ average beamforming gain is 
achievable by the proposed PDF method.

{\subsection{Extension to Multi-User Hybird 
Beamforming}\label{sec:3.4}  
In this section, we extend the proposed PDF method to multi-user hybrid 
beamforming systems in Section~\ref{sec:2}. The same DPP architecture depicted 
in 
Fig.~\ref{img:DPP} is employed for each RF chain. Specifically, recall that the 
introduction 
of TTD circuits makes the analog beamformer frequency-dependent. Therefore, we 
denote this new analog beamformer as $\mb{F}_{m} = [\mb{f}_{0,m}, 
\mb{f}_{1,m}, \cdots, \mb{f}_{U-1,m}]$, where each $\mb{f}_{u,m}$ satisfies the 
circuit restriction of DPP:
\begin{align}\label{eq:fum}
	\left\{\begin{array}{l}
		\mb{f}_{u,m} = [\mb{f}_{u,m}^{(0)^T}, \mb{f}_{u,m}^{(1)^T},\cdots, 
		\mb{f}_{u,m}^{(K-1)^T}]^T, \\
		\mb{f}_{u,m}^{(k)} = \frac{e^{-j k_m 
				r_{u,k}'}}{\sqrt{N}}[e^{j\pi\delta_P^{(0)}\beta_{u,k}'},\cdots, 
		e^{j\pi\delta_P^{(P-1)}\beta_{u,k}'}]^T.
	\end{array}
	\right.
\end{align} 
There, $r_{u,k}'$ and $\beta_{u,k}'$ refer to the adjustable distance parameter 
of the $k^{\rm th}$ TTD element and the adjustable phase shift parameter of the 
$k^{\rm 
th}$ sub-array connected to $u^{\rm th}$ RF chain, respectively. The basic idea 
of PDF algorithm is to align $\mb{f}_{u,m}$ with $\mb{h}_{u,m}$ user-by-user in 
the 
analog domain and eliminate the inter-user interference by the digital 
precoder. A step-by-step algorithm procedure is summarized in \textbf{Algorithm 
1}.

	\begin{algorithm*}[htb]
	\caption{$\!\!$: The proposed PDF algorithm}
	\label{alg:1}
	\begin{algorithmic}[1]
		\REQUIRE ~~\\
		Channel matrix $\mb{H}_m$; the user locations $\{(r_u, 
		\theta_u)\}|_{u 
		= 0}^{U-1}$; the total 
		transmit power $\rho$\\
		\ENSURE ~~\\
		The digital beamformer $\mb{D}_{m}$ and the analog beamformer 
		$\mb{F}_{m}$
		\\
		\FOR {$u \in \{0,1,\cdots,U-1\}$}
		\FOR {$k \in \{0,1,\cdots,K-1\}$}
		\STATE 
		Determine the distance parameter of the $k^{\rm th}$ TTD element: 
		$r_{u,k}' \leftarrow -r_{u,k} = 
		-\sqrt{r_u^2 + (\delta_K^{(k)}Pd)^2 - 
		2\delta_K^{(k)}Pdr_u\sin\theta_u}$ \\
			\STATE  Determine the phase shift of the $k^{\rm th}$ sub-array: 
$\beta_{u,k}' \leftarrow -\sin\theta_{u,k} = 
-\frac{r_u\sin\theta_u - 
			\delta_K^{(k)}Pd}{ \sqrt{r_u^2 + (\delta_K^{(k)}Pd)^2 - 
				2\delta_K^{(k)}Pdr_u\sin\theta_u}}$
		\STATE 
		Shift $r_{u,k}'$ by $L = \min_k \{r_{u,k}'\}$ to make them positive: 
		$r_{u,k}' \leftarrow L + r_{u,k}'$
		\ENDFOR
		\STATE
		Build the beamforming vector connected to the $u^{\rm th}$ RF chain 
		using \eqref{eq:fum}
		\ENDFOR
		\STATE 
		Build the analog beamformer: $\mb{F}_{m} \leftarrow [\mb{f}_{0,m}, 
		\mb{f}_{1,m}, \cdots, \mb{f}_{U-1,m}]$
		\STATE 
		Calculate the digital beamformer by ZF: $\mb{D}_{m} \leftarrow 
		\mb{F}_{m}^H\mb{H}_m^H(\mb{H}_m\mb{F}_{m}\mb{F}_{m}^H\mb{H}_m^H)^{-1}$
 		\STATE 
		 Normalize the digital beamformer to power $\rho$: $\mb{D}_{m} 
		 \leftarrow   \frac{\sqrt{\rho}\mb{D}_{m}}{\| 
		 \mb{F}_{m}\mb{D}_{m}\|_F} 
		 $
		\RETURN ${\mb{F}_{m}} $ and $\mb{D}_{m}$.
	\end{algorithmic}
\end{algorithm*}
Specifically, the analog beamformer is first determined in steps $1\sim9$. 
According to (\ref{eq:beta_k}) and (\ref{eq:t}), to align the $u^{\rm th}$ 
beamforming vector $\mb{f}_{u,m}$ with the $u^{\rm th}$ user, $r_{u,k}'$ and 
$\beta_{u,k}'$ are determined by $ - r_{u,k}$ and $-\sin\theta_{u,k}$, 
respectively. Here, $r_{u,k} = \sqrt{r_u^2 + (\delta_K^{(k)}Pd)^2 - 
	2\delta_K^{(k)}Pdr_u\sin\theta_u}$ and $\beta_{u,k}'=-\sin\theta_{u,k} = 
	-\frac{y_u - \delta_K^{(k)}Pd}{r_{u,k}} = 
-\frac{r_u\sin\theta_u - 
\delta_K^{(k)}Pd}{ \sqrt{r_u^2 + (\delta_K^{(k)}Pd)^2 - 
	2\delta_K^{(k)}Pdr_u\sin\theta_u}}$, which leads to steps $3\sim4$. Then, 
	according to \textbf{Lemma \ref{lemma:1}}, $L = \min_k\{r_{u,k}'\}$ can be 
	used to shift the distance parameters to make sure $r_{u,k}' \ge 0$, which 
	is reflected in step 5. Last, the analog beamformer $\mb{F}_m$ per 
	sub-carrier is constructed one-by-one in step $9$. As for the 
	digital precoder $\mb{D}_m$, it is built on the zero-forcing (ZF) rule to 
	eliminate 
	inter-user interference. Besides, the power of $\mb{D}_{m}$ is normalized 
	to $\rho$ to guarantee the power constraint per subcarrier, which completes 
	the PDF algorithm. 

}

 \section{Effective Rayleigh Distance}\label{sec:4}
 
	In the previous section, we assume that the user is positioned within the 
	near-field 
	range of the entire array while being in the far-field range of each 
	sub-array. 
	Therefore, it is essential to accurately identify the near-field ranges of 
	a sub-array 
	and the entire array. Traditionally, the classical Rayleigh distance is 
	employed as a standard for quantifying the near-field range. 
However, our experiments show that the Rayleigh distance overestimates 
the actual near-field range. 
For example, when the array aperture is $D = 0.384\:{\rm m}$ and the carrier 
is 
$f_c = 100$ GHz, the Rayleigh distance is around 98\:m. Yet, the far-field 
wideband beamforming method \cite{DPP_Tan2019} only exhibits a noticeable 
beamforming gain loss when the distance is less than $30\:{\rm m}$. This fact 
implies 
that classical Rayleigh distance overestimates the near-field range when 
evaluating channel capacity. 
This result is attributed to the fact that classical Rayleigh distance is 
derived by evaluating 
the  
largest phase error between planar wave and spherical wave, which does not 
directly affect the transmission rate. By contrast, 
the near-field effect has a directly impact on beamforming gain, which in 
turn plays a pivotal role in determining transmission rates. Therefore, it 
becomes apparent that a more precise metric for defining the near-field range 
in terms of
beamforming gain is required.
 	
 	Specifically, we first introduce the derivation of the classical Rayleigh 
 	distance by the evaluation of phase error. For ease of discussion, we 
 	only consider an arbitrary frequency $f_m$ and an arbitrary user. 
 	Therefore,  the subscript $m$ and $u$ is omitted in this section. 
 	Denote $\mb{h}(r, \theta)$ as the near-field channel as a function of user 
 	location $(r, \theta)$, of which
 	the 
 	$n^{\rm th}$ entry  is given by $[\mb{h}(r, \theta)]_n = g e^{-jk 
 	r^{(n)}} 
 	= g 
 	e^{-jk(r^2 + (\delta_N^{(n)}d)^2 - 2\delta_N^{(n)}dr\sin\theta)^{1/2}}$, 
 	where $g$ denotes the path loss and $k = \frac{2\pi f}{c}$ is the 
 	wavenumber. 
 	By utilizing the far-field approximation in (\ref{eq:FF2NF}), the far-field 
 	channel is expressed as $[\mb{h}_{\text{far}}(r, \theta)]_n =  
 	[\mb{h}(+\infty, \theta)]_n = g e^{-jk(r - 
 	\delta_N^{(n)}d\sin\theta)}$. Consequently, the  phase error between 
 	$[\mb{h}(r, \theta, f_c)]_n$ and $[\mb{h}_{\text{far}}(r, \theta, f_c)]_n$ 
 	is defined as \begin{align}
 		E_n(r, \theta) &= \left| \angle{[\mb{h}(r, \theta)]_n} - 
 		\angle{[\mb{h}_{\text{far}}(r, \theta)]_n} \right| \notag\\&= 
 		|kr^{(n)} - 
 		k (r - \delta_N^{(n)}d\sin\theta)|.
 		\end{align}
 	Subsequently, the definition of Rayleigh distance is as follows 
 	\cite{fresnel_Selvan2017}: if the distance $r$ from user to BS 
 	exceeds the Rayleigh distance $R$, then the largest phase error 
 	$E(r) = \max_{n, \theta}{E_n(r, \theta)}$ is no more than $\frac{\pi}{8}$.
 	That is to say, once the largest phase error $E(r)$ surpasses
 	$\frac{\pi}{8}$, the user is situated in the near-field region.
 	To derive the close-form expression of $E(r)$, the second-order Taylor 
 	expansion $(1 + x)^{\frac{1}{2}} \approx 1 + \frac{1}{2}x - \frac{1}{8}x^2$ 
 	is commonly used \cite{NearCE_Cui21} to approximate the distance 
 	$r^{(n)}$ as 
 	\begin{align} \label{eq:app}
 	r^{(n)} &= \sqrt{r^2 - 2r\delta_N^{(n)}d\sin\theta + 
 	(\delta_N^{(n)}d)^2} 
 	\notag\\ 
 	&\approx  r - \delta_N^{(n)}d\sin\theta + 
 	\frac{(\delta_N^{(n)}d)^2\cos^2\theta}{2r}.
 	\end{align}
    In the field of microwave and antenna, the approximation \eqref{eq:app} is 
    known as 
    the Fresnel approximation~\cite{fresnel_Sherman1962}. 
 	Next, the phase error can be approximated as $E_n(r, \theta) \approx k_c 
 	\frac{(\delta_N^{(n)}d)^2\cos^2\theta}{2r}$. Given that $\cos^2\theta \le 
 	1$ and $\delta_N^{(n)} = n 
 	- 
 	\frac{N - 1}{2}$, we 
 	have \begin{align}
 		E(r) = \max_{n, \theta}{E_n(r, \theta)} \approx k \frac{(0.5({N - 
 		1})d)^2}{2r} = \frac{D^2  
 	\pi}{4r\lambda}.
 	\end{align}  
 In order for $E(r)\le \frac{\pi}{8}$, it is necessary to make $r \ge 
 \frac{2D^2}{\lambda}$.
 	Consequently, the Rayleigh distance is given by
 	\begin{align}\label{eq:RD}R \approx 
 	\frac{2D^2}{\lambda}.
 \end{align}
 	
 	On the other hand, we define a new effective Rayleigh distance via the 
 	evaluation of  beamforming gain loss.
 	To elaborate, the normalized coherence between the channel $\mb{h}(r, 
 	\theta)$ and its far-field approximation 
 	$\mb{h}_{\text{far}}(r, \theta)$ is 
 	characterized by
 	\begin{align}\mu(r, \theta) = \frac{1}{|g|^2 N }|\mb{h}^{H}(r, 
 	\theta)\mb{h}_{\text{far}}(r, \theta)|.\end{align}
 	The coherence $\mu(r, \theta)$ equivalents to the achievable beamforming 
 	gain at frequency $f$ when the BS utilizes the far-field beamforming 
 	vector 
 	$\mb{f} = \frac{1}{|g| \sqrt{N} }\mb{h}_{\text{far}}^*(r, \theta)$ 
 	to serve a user located at $(r, \theta)$. Clearly, this beamforming 
 	gain would gradually decline when the user is moving close to BS and the 
 	near-field effect becomes remarkable. When the beamforming gain loss, 
 	denoted as $1 - 
 	\mu(r, \theta)$, exceeds a predefined threshold $\Delta$, it indicates that
 	the 
 	user has entered the near-field region. Consequently, the boundary 
 	$R_{\text{eff}}$, where $1 - 
 	\mu(R_{\text{eff}}, \theta)$ exactly equals to $\Delta$, is 
 	defined as the effective Rayleigh distance. Notably, the direct influence 
 	of 
 	beamforming gain $\mu(r, \theta)$ on the received signal power makes
 	$R_{\text{eff}}$ a more accurate metric for characterizing the 
 	near-field range in communication systems. \textbf{Lemma \ref{lemma:3}} 
 	gives out the 
 	close-form expression of effective Rayleigh distance $R_{\text{eff}}$.
 	 \begin{lemma} \label{lemma:3}
 	  	We define the effective Rayleigh distance $R_{\text{eff}}$ such that 
 	  	the inequality $1 - 
 	  	\mu(r, \theta) \ge \Delta$ always holds for $0 < r \le R_{\text{eff}}$. 
 	  	Then,  the value of $R_{\text{eff}}$ is given by
 	  	\begin{align}\label{eq:ERD}
 	  	R_{\text{eff}} \approx C_\Delta \cos^2\theta \frac{2D^2}{\lambda},
 	  	\end{align}
   		where $C_\Delta = \frac{1}{4\beta_\Delta^2}$ and $\beta_\Delta$ is the 
   		solution of the equation $\frac{1}{\beta_\Delta}| 
   		\int_{0}^{\beta_\Delta}e^{-j\frac{1}{2}\pi t^2} 
   		\text{d}t 
   		| = \Delta$.
 	  	
 	\end{lemma}

 	\begin{IEEEproof}
 	To obtain the value of the effective Rayleigh distance, we need to derive 
 	the close-form expression of $\mu(r, \theta)$. Based on the second-order 
 	Taylor expansion in (\ref{eq:app}), $\mu(r, \theta)$ can be expressed as 
 	\begin{align}\label{eq:f}
	\mu(r, \theta) & \approx \frac{1}{N} \left| \sum_{n = 0}^{N - 
	1}e^{-j\pi\frac{(\delta_N^{(n)}d)^2\cos^2\theta}{\lambda r}} \right| 
	\notag\\
	& = \frac{1}{N} \left| \sum_{m = - \frac{1}{2} + \frac{1}{2N}}^{\frac{1}{2} 
	- \frac{1}{2N}}e^{-j\pi\frac{m^2(Nd)^2\cos^2\theta}{\lambda r}} \right|.
\end{align}
 	Notice that the operator $\sum_{m = - \frac{1}{2} + 
 	\frac{1}{2N}}^{\frac{1}{2} - \frac{1}{2N}}$ performs the summation over 
 	$m = - \frac{1}{2} + \frac{1}{2N}, - \frac{1}{2} + \frac{3}{2N}, - 
 	\frac{1}{2} + \frac{5}{2N}, \cdots, \frac{1}{2} - \frac{1}{2N}$. 
 	Let's define $\zeta = \frac{N^2d^2\cos^2\theta}{\lambda r}$ for brevity.
Since the number of antennas $N$ is quite large, (\ref{eq:f}) can be 
represented 
in an integral form as 
 	\begin{align}\label{eq:f2}
 	\mu(r, \theta) =  | \int_{-1/2}^{1/2}e^{-j\pi m^2 \zeta} \text{d}m 
 	| 
 	= 2 | \int_{0}^{1/2}e^{-j\pi m^2 \zeta} \text{d}m |.
 	\end{align}
 	Additionally, we introduce the variable transformation: $\frac{1}{2}t^2 = 
 	m^2\zeta$. Then, $\mu(r, \theta)$ can be 
 	rewritten as
 	\begin{align}\label{eq:ap2}
 	\mu(r, \theta) =  \frac{2}{\sqrt{2\zeta}}\left| 
 	\int_{0}^{\frac{\sqrt{2\zeta}}{2}}e^{-j\frac{1}{2}\pi t^2} \text{d}t 
 	\right| = G(\beta),
 	\end{align}
 	where $G(\beta) = \left| \int_{0}^{\beta}e^{-j\frac{1}{2}\pi t^2} \text{d}t 
 	\right| / \beta$ and $\beta = \frac{\sqrt{2\zeta}}{2} = 
 	\sqrt{\frac{N^2d^2\cos^2\theta}{2\lambda  r}} = 
 	\sqrt{\frac{D^2\cos^2\theta}{2\lambda  r}}$. It is clear from 
 	(\ref{eq:ap2}) that the coherence heavily relies  on the characteristics of 
 	the function $G(\beta)$. Fortunately, $G(\beta)$ does not contain any 
 	parameters, allowing us to obtain its numerical result via offline 
 	integration.
 	\begin{figure}	
 		\centering
 		\includegraphics[width=3.5in]{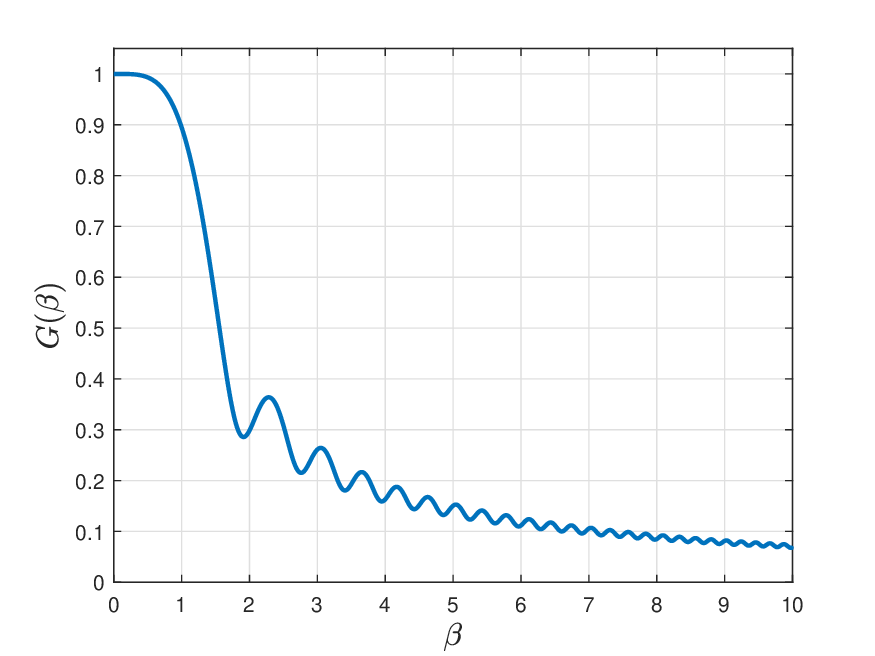}
 		\caption{ The numerical results of $G(\beta)$
 		}\label{img:gx}
 		\vspace*{-1em}
 	\end{figure}
 	
 	As illustrated in Fig. \ref{img:gx}, the function  $G(\beta)$ shows a 
 	significant 
 	downward trend w.r.t $\beta$\footnote{ The function $G(\beta)$ 
 	is applied in our another paper~\cite{LDMA_Wu2023} as well to evaluate the 
 	quasi-orthogonality of 
 	near-field channels, while $G(\beta)$ is used to 
 	derive the 
 	effective Rayleigh distance in this paper.
 }. Therefore, to make the 
 	beamforming gain loss 
 	$1 - \mu(r, 
 	\theta) = 
 	1 - G(\beta)$ larger than the threshold $\Delta$, we need $\beta \ge 
 	\beta_\Delta$, where $G(\beta_\Delta) = 1 - \Delta$. 
 	
 	In the end, due to the relationship $\beta =  
 	\sqrt{\frac{D^2\cos^2\theta}{2\lambda  r}}$, 
 	the near-field region is determined by 
 	$r < \frac{1}{4\beta_\Delta^2}\cos^2\theta 
 	\frac{2D^2}{\lambda }$, giving rise to the result $R_{\text{eff}} = 
 	\frac{1}{4\beta_\Delta^2}\cos^2\theta 
 		\frac{2D^2}{\lambda }$. 
 	\end{IEEEproof}
 	
\textbf{Lemma \ref{lemma:3}} offers a comprehensive approach to compute 
$R_{\text{eff}}$. 
To illustrate, let's consider a simple example where  $\Delta = 5\%$. We could 
solve the equation 
$\frac{1}{\beta_\Delta}| 
\int_{0}^{\beta_\Delta}e^{-j\frac{1}{2}\pi t^2} 
\text{d}t 
| = 0.05$ via the Newton method and obtain $\beta_\Delta = 0.8257$. Hence, the 
effective Rayleigh distance is evaluated as $R_{\text{eff}} = 0.367\cos^2\theta 
\frac{2D^2}{\lambda}$. 

It is evident from (\ref{eq:RD}) and (\ref{eq:ERD}) that effective Rayleigh 
distance needs two more 
variables compared to Rayleigh distance, i.e., the constant $C_\Delta$ related 
to beamforming gain loss and the angle of arrival $\theta$.
These two variables enable effective Rayleigh distance to accurately capture  
where far-field beamforming are not applicable, and thus make it a more 
accurate metric for quantifying near-field region. 
 	In Section \ref{sec:5}, the accuracy of effective Rayleigh distance will be 
 	verified through simulation.

 \subsection{Discussion on the Fresnel Approximation}\label{sec:4-1} 
In deriving the effective Rayleigh distance, the Fresnel 
approximation~\eqref{eq:app} is employed. As indicated in 
\cite{fresnel_Selvan2017}, the approximation \eqref{eq:app} is accurate 
when the 
distance $r$ is larger than the ``Fresnel distance" 
$0.5\sqrt{\frac{D^3}{\lambda_c}}$. To validate the rationality 
of the Fresnel approximation, we would like to show that the effective Rayleigh 
distance 
$0.367\cos^2\theta\frac{2D^2}{\lambda_c}$, with $\Delta = 5\%$, is much larger 
than the Fresnel 
distance. 
Take it into account that the section range of a typical cell is 
around $\frac{2\pi}{3}$, thus $\theta$ is restricted between $-\frac{\pi}{3}$ 
and 
$\frac{\pi}{3}$. 
Accordingly, $0.367\cos^2\theta\frac{2D^2}{\lambda_c} \ge 0.5 
\sqrt{\frac{D}{\lambda_c}}$ is equivalent to $N > 15.8490$. Given that an 
extremely large antenna array have hundreds or thousands of 
antennas, which is greatly larger than $15.8490$, 
the effective Rayleigh distance is much longer than the Fresnel distance,  
resulting in the accuracy of approximation 
\eqref{eq:app}.

 \subsection{Discussion on the Piecewise-Far-Field Approximation} 
 The effective Rayleigh distance is capable of verifying the accuracy of the 
 piecewise-far-field approximation as well. Recall that the user should 
 locate in the far-field region of each sub-array, i.e., $r_k$ must 
 be larger than $0.367\cos^2\theta_k \frac{2(P-1)^2d^2}{\lambda_c}$ with 
 $\Delta = 5\%$. 
 Take a small sub-array configuration as an example: $P = 32$ and $f_c = 
 100\:{\rm 
 	GHz}$. The effective 
 	Rayleigh distance per sub-array is 
 upper bounded by 
 $0.367\cos^2\theta_k\frac{2(P-1)^2d^2}{\lambda_c}\le0.367\frac{2(P-1)^2d^2}{\lambda_c}
  = 0.5286\:{\rm 
 	m}$. In this context, as long as the user-to-sub-array distance $r_{k}$ is  
 larger than $0.5286\:{\rm m}$, a common situation in mobile communications, 
 each sub-array's channel can be precisely 
 modeled as far-field. Therefore, under a small sub-array configuration, the 
 piecewise-far-field approximation is accurate. 
 
 	 \subsection{Discussion on the Number of Antennas per 
 	 Sub-Array}\label{sec:4-2} 
 	In this sub-section, by combining \textbf{Lemma \ref{lemma:1}}, 
 	\textbf{Lemma \ref{lemma:2}}, and 
 	\textbf{Lemma \ref{lemma:3}}, the value of the essential parameter $P$, the 
 	number 
 	of antennas per
 	sub-array, is designed. 
 	
 	The value of  $P$ needs to meet three key requirements. First, 
 	as stated in 
 	\textbf{Lemma \ref{lemma:1}}, $ | \epsilon_m | \le \frac{2}{P}$ holds for 
 	$\forall m$. Owing to the fact that $\max | \epsilon_m | = \frac{B}{2f_c}$, 
 	we 
 	have $P \le 
 	\frac{4f_c}{B}$. 
 	
 	Next, we made an assumption that the user is located in 
 	the far-field region of each 
 	sub-array. We evaluate the effective Rayleigh distance at the center 
 	frequency $f_c$ with a wavelength $\lambda_c$. 
 	Suppose the user's activity 
 	range is $r \in [\rho_{l}, 
 	\rho_{h}]$ and $\theta \in [-\theta_{h}, \theta_{h}]$, 
 	where 
 	$\rho_{l}$ and $\rho_{h}$ can be regarded as 
 	the least allowable distance from user to BS and the cell radius, and 
 	$\theta_{h}$ refers to the sector range of a cell. Applying 
 	\textbf{Lemma \ref{lemma:3}}, the effective 
 	Rayleigh distance of a sub-array is 
 	$C_\Delta\cos^2\theta\frac{2(Pd)^2}{\lambda_c} = \frac{1}{2}C_\Delta 
 	\cos^2\theta 
 	P^2\lambda_c \le
 	 \frac{1}{2}C_\Delta P^2\cos^2\theta_h\lambda_c^2$. 
 	Then, to keep the user consistently outside the region bounded 
 	by 
 	$\frac{1}{2}C_\Delta 
 	\cos^2\theta 
 	P^2\lambda_c$, we can let 
 	$\rho_l 
 	\ge \frac{1}{2}C_\Delta \cos^2\theta_h P^2\lambda_c$ and arrive at the 
 	condition $P 
 	\le 
 	\sqrt{\frac{2\rho_{l}}{C_\Delta \cos^2\theta_h\lambda_c}}$.  
 	
	Finally, in order to guarantee the performance of PDF method, we would like 
	to design $P$ such that the least average beamforming gain in 
	\textbf{Corollary 1} is greater than a predefined threshold $\delta$. This 
	requirement can be formulated as follows:
	\begin{align}
		\min_{r,\theta} G &= 1 - \gamma(B, f_c, P)\max_{r,\theta}\xi(r, \theta, 
		D) \ge \delta. \notag \\
		&\Rightarrow \Xi_P(\frac{B}{2f_c}) \ge 1 - \frac{3(1 - 
		\delta)}{\max_{r, 
		\theta} \xi(r, \theta, D)}
	\end{align} 	
	Notice that $\max_{r,\theta}\xi(r, \theta, D) = \max_{r} \xi(r, \theta_h, 
	D) $ given that $\xi(r, \theta, D)$ is a decreasing function w.r.t 
	$|\theta|$. 
	Furthermore,  
	we can utilize the gradient ascend method to solve $\max_{r} \xi(r, 
	\theta_h, D)$, and thereafter employ the Newton method to attain 
	$P_\delta$ 
	from the 
	equation $ \Xi_{P_{\delta}}(\frac{B}{2f_c}) = 1 - \frac{3(1 - 
		\delta)}{\max_{r} \xi(r, \theta_h, D)} $. 
	Finally, taking it into consideration the monotonic decreasing property of 
	function $ \Xi_{P}(\frac{B}{2f_c})$ w.r.t $P$, we can draw the conclusion 
$P \le 
	P_\delta$.
	
 	As a result, applying the three requirements above, the number of 
 	antennas per sub-array $P$ should satisfy 
 	\begin{align} \label{eq:req}
 		P \le \min\left\{ \frac{4f_c}{B} ,  \sqrt{\frac{2\rho_{l}}{C_\Delta 
 		 \cos^2\theta_h\lambda_c}}, P_\delta
 		\right\}.
 	\end{align}
 For instance, considering the following parameters: $N = 400$, $B = 5$ GHz, 
 $f_c = 
 100$ GHz, $\rho_{l} = 1$ m, $\theta_{h} = \frac{\pi}{3}$, $\Delta = 5\%$, and 
 $\delta = 90\%$, we have $\frac{4f_c}{B} = 80$, $ 
 \sqrt{\frac{2\rho_{l}}{C_\Delta 
 		\cos^2\theta_{h} \lambda_c}} \approx 
 	43$ and $P_{\delta} \approx 42 $. Therefore, the number of antennas per 
 	sub-array can be $P = 40$, 
 	meaning that 
 	a $400$-antenna array just needs $K = \frac{400}{40} = 10$  TTDs  to 
  alleviate the near-field beam split effect.
 	 Moreover, the lower bound of $P$ is exactly 
 	1, because a 
 	reduced $P$ leads to an increased deployment of TTD units, enabling the PDF 
 	method to achieve more flexible frequency-dependent beamforming. In this 
 	context, the beamforming gain is increasingly improved with the reduction 
 	of $P$.

 		\section{Simulation Results} \label{sec:5}
 		\begin{table}[]
 			\centering
 			\caption{ System Configurations}
 			\begin{tabular}{|c|c|}
 				\hline
 				The number of the BS antennas $N$ & 256 \\ \hline
 				The number of Users $U$ & 1, 4 \\ \hline
 				The center frequency $f_c$                 & 100 
 				GHz                    \\ \hline
 				The bandwidth $B$                          & 5 GHz \\ \hline
 				The number of subcarriers $M$              & 
 				256                        \\ \hline 
 				The user's activity range $[\rho_l, \rho_h]$ & [1 m, 100 m] \\ 
 				\hline 
 				The sector range of a cell $\theta_h$ & $\pi/3$ \\ 
 				\hline 
 				Threshold parameter $\Delta$ & 5\% \\\hline
 				Threshold parameter $\delta$ & 90\% \\\hline 
 				The number of a sub-array's antennas $P$ & 32 \\\hline 
 				The number of TTDs $K$ & 8\\\hline
 			\end{tabular}
 		\end{table}
 		
 	In this section, numerical results are provided to demonstrate the 
 	performance 
 	of the proposed PDF method and the accuracy of effective Rayleigh distance.
 	The default simulation parameters are presented in Table I unless 
 	particularly 
 	specified.

	\subsection{Beamforming Gain}\label{sec:5-1}
	\begin{figure}
		\centering
		\subfigure[Traditional narrowband beamfocusing 
		\cite{THzbeam_Headland2018}]
		{\includegraphics[width=3.5in]{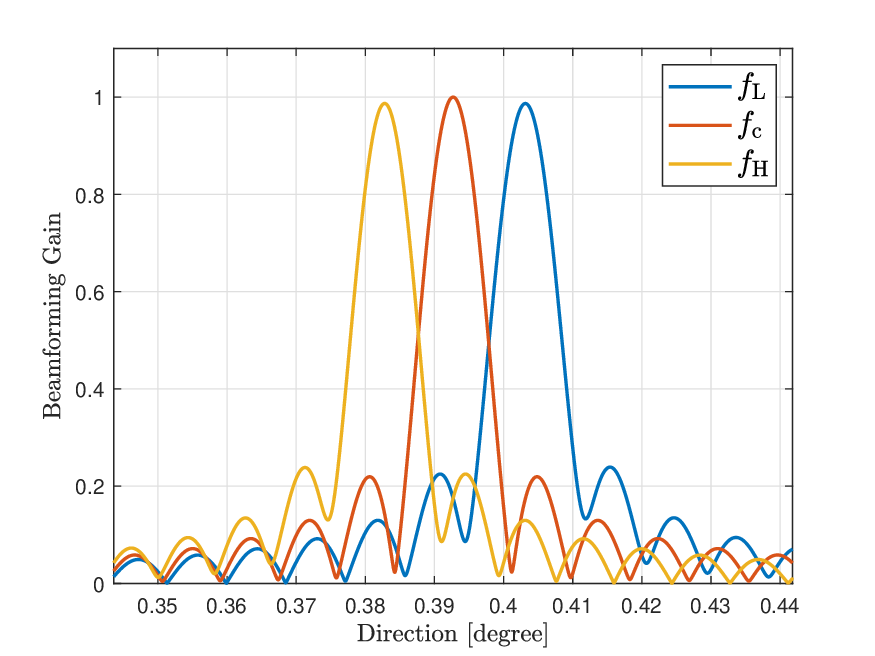}}
		 
		\subfigure[Proposed PDF method]
		{\includegraphics[width=3.5in]{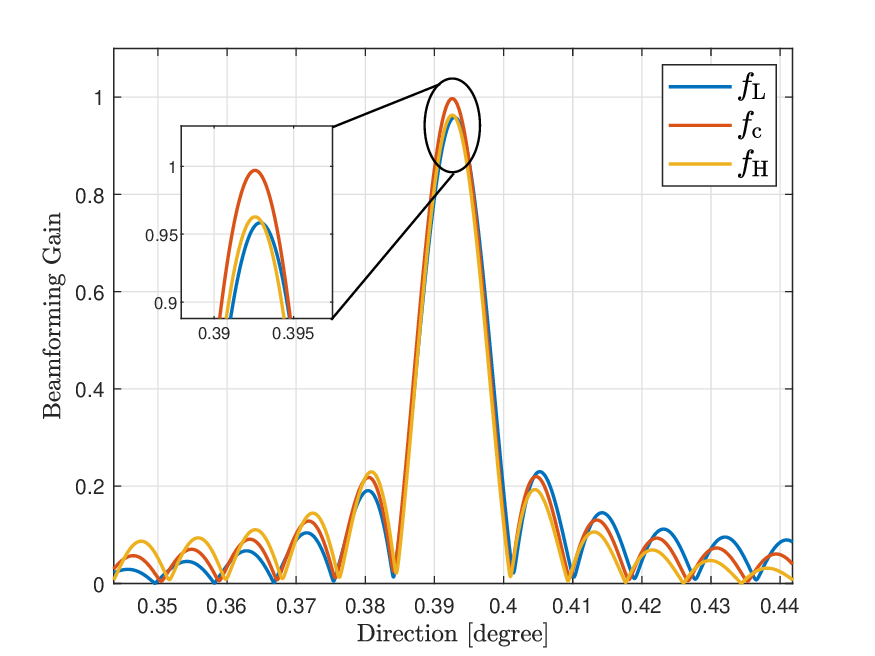}}
		\\ 
		\centering
		\caption{Beamforming gain per sub-carrier w.r.t direction.} 
	\label{img:BG}
	\end{figure}
	
	We begin with comparing the beamforming gain performance for single-user 
	scenarios, i.e., $U = 1$.
	
	Fig. \ref{img:BG} presents the beamforming gain performance for different 
	sub-carriers as a function of direction. Here, the user is located at $(r, 
	\theta) = (2\:\text{m}, \frac{\pi}{8})$, and we evaluate the beamforming 
	gain for various frequencies and physical directions. 
	Fig. \ref{img:BG}(a) showcases the beamforming gain achieved by 
	traditional 
	near-field narrowband beamfocusing method \cite{THzbeam_Headland2018}, 
	while 
	Fig. \ref{img:BG}(b) presents the results of the proposed PDF method. 
	Let  $f_{\text{L}}$, $f_{\text{c}}$, and $f_{\text{H}}$ be the 
	lowest, the center, and the highest frequency, respectively.  
	For the traditional narrowband beamfocusing method , 
	the near-field beam split effect causes the beams at $f_{\text{L}}$, 
	$f_{\text{c}}$, and  $f_{\text{H}}$ to be focused on different locations, 
	leading to a significant beamforming gain loss at $f_{\text{L}}$ and 
	$f_{\text{H}}$.
	However, as illustrated in Fig. \ref{img:BG} (b), the proposed PDF 
	method can 
	effectively focus the energy of the beams at $f_\text{L}$, $f_\text{c}$, 
	and $f_\text{H}$ on the desired user location. Besides, more than 95\% 
	beamforming gain on the user location is achieved for 
	$f_\text{L}$ and $f_\text{H}$. 
	Therefore, the proposed PDF method is able to effectively mitigate the 
	near-field beam split effect.
	
		\begin{figure}
		\centering

		\includegraphics[width=3.5in]{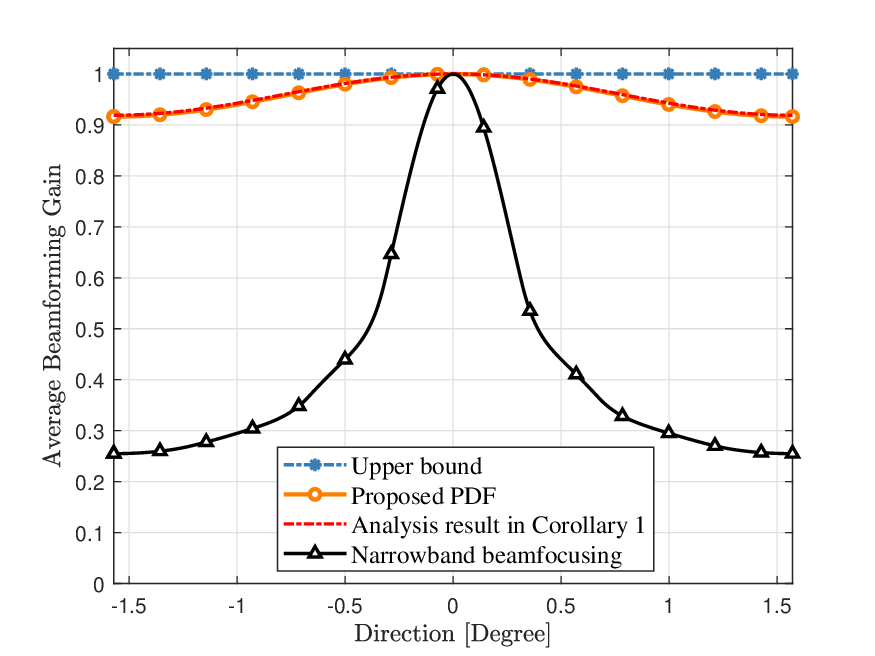}
		\\ 
		\centering
		\caption{Average beamforming gain w.r.t direction $\theta$.} 
		\label{img:ABG_theta}
	\end{figure}
	Fig. \ref{img:ABG_theta} illustrates the average beamforming gain 
	performance w.r.t direction $\theta$. In this context, the distance $r$ is
	fixed as 10 m
	and the direction $\theta$ spans from $-\theta_h$ to $\theta_h$. We 
	can observe that the analysis result (\ref{eq:C2}) for average beamforming 
	gain is quite close to the real average beamforming gain achieved by the 
	PDF 
	method. With the increment of $|\theta|$, the average 
	beamforming gain achieved by both the PDF method and the narrowband 
	beamfocusing method declines. This is attributed to the fact that the beam 
	split effect becomes more significant with larger $|\theta|$. 
	Nevertheless, our PDF method could remain more than $\delta = 90 \%$ 
	average beamforming gain over $\theta 
	\in [-\theta_h, \theta_h]$, which is 
	consistent with our discussion on the number of a sub-array's numbers in 
	section \ref{sec:4-2}.

		\begin{figure}
	\centering
	
	\includegraphics[width=3.5in]{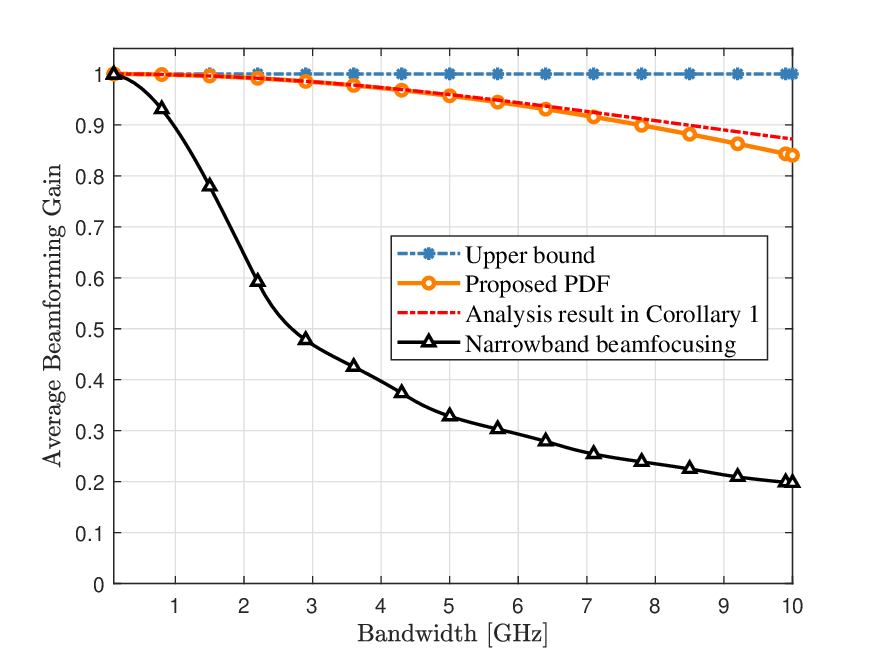}
	\\ 
	\centering
	\caption{Average beamforming gain performance w.r.t bandwidth $B$.} 
	\label{img:ABG_B}	
\end{figure}
	
Moreover, Fig. \ref{img:ABG_B} shows the average beamforming gain 
performance w.r.t bandwidth $B$. In this simulation, the user is located at 
$(r, \theta) = (10\:\text{m}, \frac{\pi}{4})$, and the bandwidth increases 
from 100 MHz to 10 GHz. It is clear from Fig. \ref{img:ABG_B} that our PDF 
method could remarkably enhance the near-field beamforming capability by 
offering near optimal average beamforming gain. For instance, when $B = 5$ GHz, 
around 3 times higher average beamforming gain is reaped by the PDF method than 
narrowband beamfocusing. Besides, it is notable that when the bandwidth is 
around 10 GHz, our analysis results slightly differ from the real performance 
of PDF. This is because, with larger bandwidth $B$, the precision of quadratic 
fitting in (\ref{eq:ApproSinc}) gets reduced, leading to an error of the
analytical beamforming gain.

	\subsection{Spectral Efficiency}\label{sec:5-2}
	
				\begin{figure}
		\centering
		\includegraphics[width=3.5in]{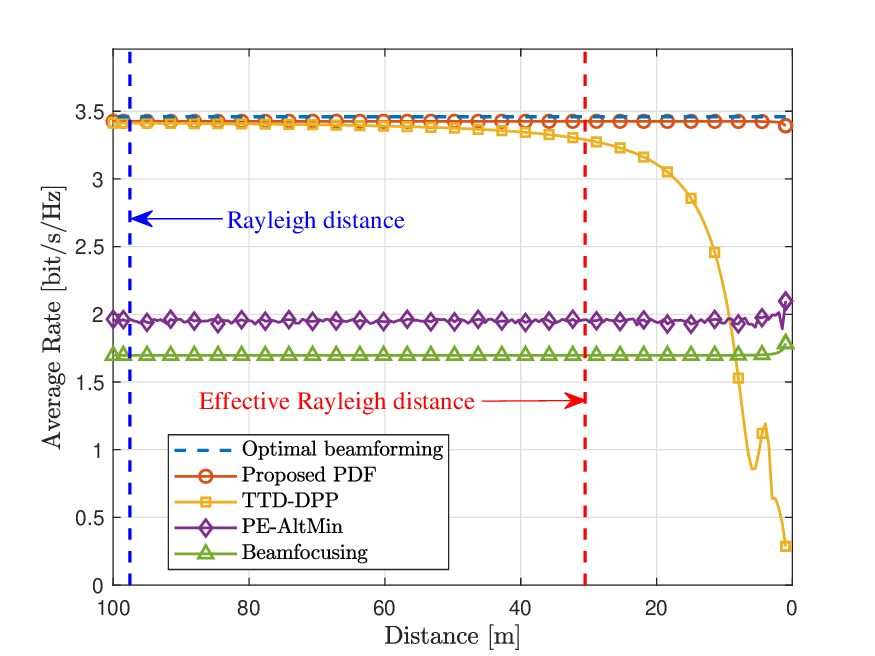}
		\\ 
		\centering
		\caption{Spectral efficiency w.r.t distance $r$.
		} 
		\label{img:SR_dis}
	\end{figure}

	In this subsection, the spectral efficiency, formulated as 
	\begin{align}	
    {\rm SE} = \frac{1}{M}\sum_{m=0}^{M-1}\sum_{u = 
    0}^{U-1}\log_2\left(1+\frac{\left|\mb{h}_{u,m}^T\mb{F}_{m} \mb{d}_{u, m}
		\right|^2}{\sum_{v\neq u}\left|\mb{h}_{u,m}^T\mb{F}_{m} \mb{d}_{v, m}
		\right|^2 + \sigma^2} \right)
	\end{align} is evaluated to compare different beamforming algorithms, where 
	$\mb{d}_{u,m}$ represents the $u^{\rm th}$ column of $\mb{D}_{u,m}$. 
	The path gains $g_m$ are generated from the Complex Gaussian distribution 
	$\mathcal{CN}(0,1)$. The signal-to-noise ratio (SNR) is defined as 
	 $\text{SNR} =\frac{\rho}{\sigma^2}$.  
	
	The compared algorithms include the narrowband 
	beamfocusing method in \cite{THzbeam_Headland2018}, the PE-AltMin algorithm 
	 for  wideband hybrid beamforming designed in \cite{MOALT_Yu2016}, and 
	the TTD-DPP algorithm tailored for solving far-field 
	beam split in \cite{DPP_Tan2019}. 
	Last, the optimal beamforming achieved by fully TTD arrays, where all phase 
	shifters in Fig.~\ref{img:layout} are replaced with TTD circuits and 
	$\mb{f}_{u,m}$ is constructed as $\mb{a}_{m}^*(r_u,\theta_u)$ in 
	Algorithm 1, is 
	employed as the performance upper bound. 

	To begin with, the spectral efficiency w.r.t the distance $r$ is 
	depicted 
	in Fig. \ref{img:SR_dis}. The number of users is set as $U = 1$.
	To explicitly illustrate the impact of the 
	near-field effect, we keep the SNR as 10 dB for different distances, 
	where 
	the large-scale fading is 
	compensated by transmit power control. 
	The user moves from  $(\rho_h, \frac{\pi}{8})$ to $(\rho_l, 
	\frac{\pi}{8})$ in a straight line. 	We can observe from Fig. 
	\ref{img:SR_dis} that our PDF method outperforms all compared beamforming 
	methods over all distances, and approaches the optimal beamforming. This 
	is attributed to the fact that our PDF method 
	can tackle the near-field effect and beam split effect 
	simultaneously. Moreover, notice that with 256 antennas at $f_c = 
	100$ GHz, the classical 
	Rayleigh distance is around 98 meters. However, as illustrated in Fig. 
	\ref{img:SR_dis}, the far-field beamforming algorithm TTD-DPP 
	\cite{DPP_Tan2019} 
	does not exhibit a notable rate loss until the distance is less than 
	30 meters, which implies that the Rayleigh distance overestimates the 
	near-field range when evaluating the communication rate. This is 
	attributed to the definition of Rayleigh distance from the phase error 
	\cite{fresnel_Selvan2017}, which has no direct influence on the spectral 
	efficiency. In contrast, since the beamforming gain makes a direct 
	influence to 
	the 
	received signal power, our newly defined effective Rayleigh distance 
	(\ref{eq:ERD}) is a 
	more accurate metric to quantify the near-field range for
	communications. Specifically, with $\theta = \pi/8$, $N = 256$, $f_c = 100$ 
	GHz, and $\Delta = 5\%$, we have $R_{\text{eff}} \approx 31$ m.
	As shown in Fig. \ref{img:SR_dis}, the spectral efficiency achieved by the 
	far-field method TTD-DPP starts declining exactly when the 
	distance is less than $R_{\text{eff}}$, which demonstrates the accuracy of 
	effective Rayleigh distance.

	\begin{figure}
	\centering
	\includegraphics[width=3.5in]{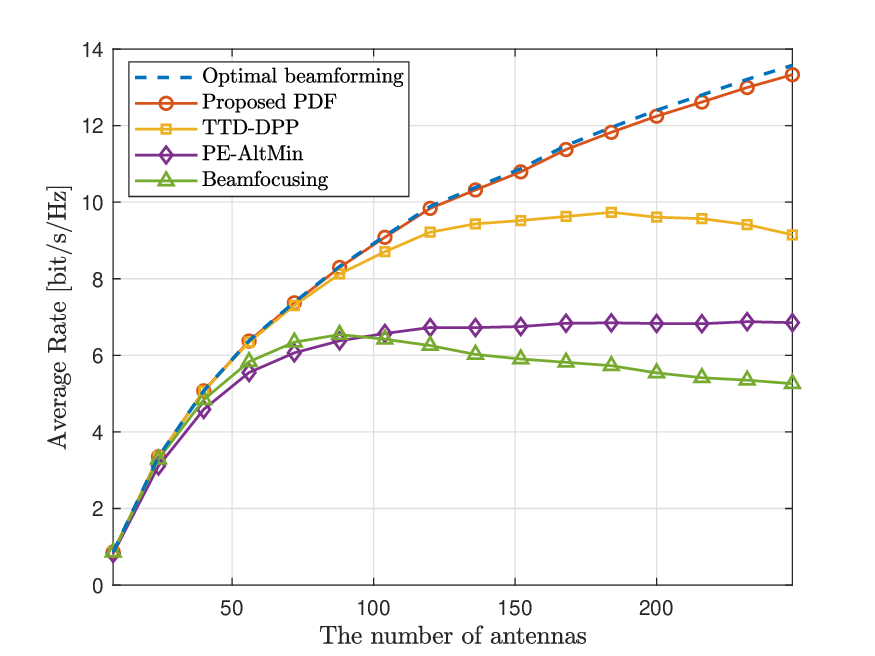}
	\\ 
	\centering
	\caption{Spectral efficiency w.r.t the number of antennas $N$.
	} 
	\label{img:SR_N}
\end{figure}

Fig. \ref{img:SR_N} evaluates the spectral efficiency w.r.t the number of 
BS antennas $N$ ranging from 8 to 256. The 
other configuration is set as follows: $U = 4$; $K = 8$; ${\rm SNR} = $ 
10~dB; the distances $\{r_u\}_{u = 0}^{U-1}$ and directions   $\{\theta_u\}_{u 
= 0}^{U-1}$
are sampled from the uniform distribution $\mathcal{U}(1\:{\rm m}, 
30\:{\rm 
m})$ and $\mathcal{U}(-\frac{\pi}{3}, \frac{\pi}{3})$, respectively. The 
spectral 
efficiency are obtained through 10000 times
Monte-Carlo simulations.  
To elaborate, Fig. \ref{img:SR_N} can be divided into three regions according 
to the number of BS antennas. 
When $N < 50$, the array aperture is small, so the user is located in the 
far-field area and the beam split effect is negligible. In this context, all
algorithms could achieve good performance. 
Next, when $50 < N < 100$, the far-field beam split effect appears, leading to 
a severe degradation to the narrowband beamfocusing and PE-AltMin methods. On 
the other hand, since TTD-DPP and PDF can effectively 
alleviate far-field beam split, both of them could remain near-optimal average 
rate performance. 
Finally, when the number of BS antennas is further increased such that $N > 
100$, the array aperture is quite large so that the near-field
beam split effect is observed. 
In this context, the TTD-DPP method tailored for far-field beam split is not 
applicable any more, while the proposed PDF method can remain a stable 
beamforming performance by mitigating the near-field beam split.


\begin{figure}
	\centering
	\includegraphics[width=3.5in]{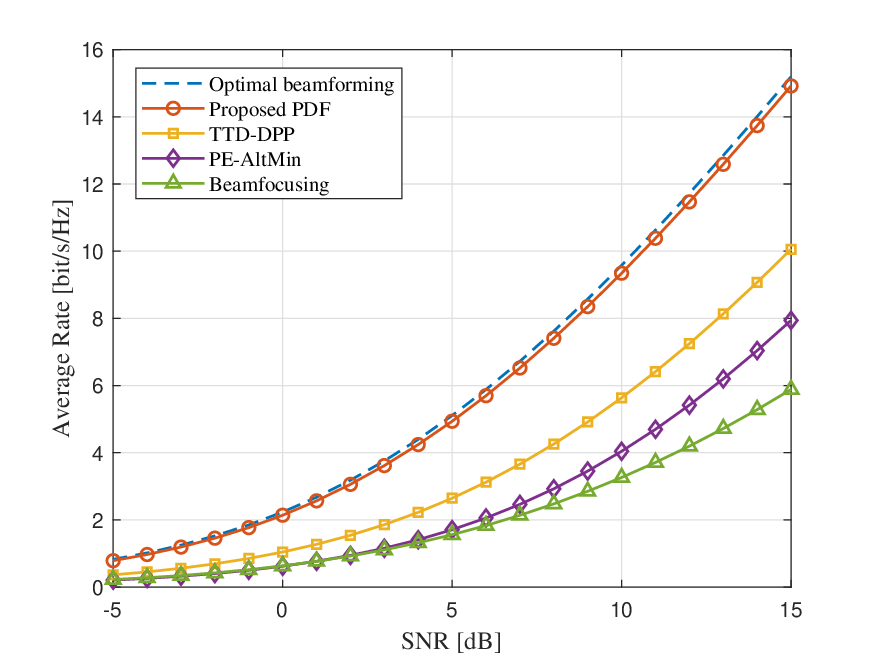}
	\\ 
	\centering
	\caption{Spectral efficiency w.r.t SNR. 
	} 
	\label{img:SR_SNR}
\end{figure}

 The influence of SNR on the spectral efficiency is investigated 
	in Fig. \ref{img:SR_SNR}, where the SNR increases 
from -5 dB to 10 dB. The other settings are as follows: $U = 4$; $r_u\sim 
\mathcal{U}(1\:{\rm m}, 
30\:{\rm 
	m})$; $\theta_u \sim \mathcal{U}(-\frac{\pi}{3}, 
\frac{\pi}{3})$. 
The spectral efficiency are obtained through 10000 times
Monte-Carlo simulations. 
With the increment of SNR, the spectral efficiency of our PDF 
method rises rapidly. When $\text{SNR} = 10\:\text{dB}$, more than 3 bit/s/Hz 
improvement in 
spectral efficiency is achievable by the PDF method compared  to the TTD-DPP 
method, 
which 
further strengthens the superiority of the PDF method.

	\subsection{Energy Efficiency}\label{sec:5-3}
	\begin{figure}
	\centering
	\includegraphics[width=3.5in]{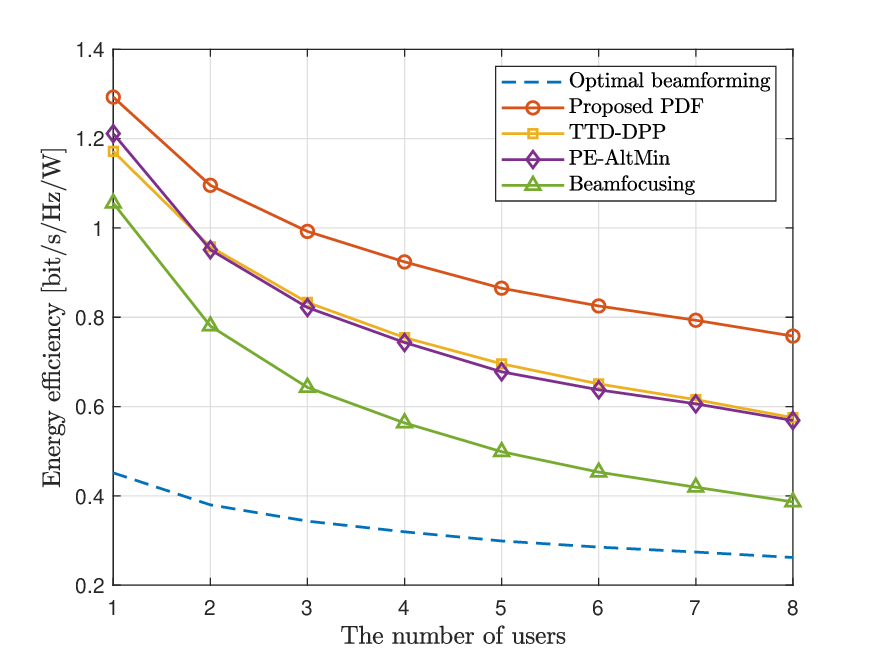}
	\\ 
	\centering
	\caption{Energy efficiency w.r.t the number of users $U$. 
	} 
	\label{img:EE_U}
\end{figure}

	Fig.~\ref{img:EE_U} provides an energy efficiency comparison when $U = 
	N_{\rm RF}$ varies from 1 to 8. The energy efficiency is defined as the 
	ratio 
	between the spectral efficiency and the total power consumed by 
	the baseband digital 
processing and analog circuits.  
The compared benchmarks include the PE-AltMin~\cite{MOALT_Yu2016} and 
narrowband beamfocusing~\cite{THzbeam_Headland2018} methods based on 
conventional hybrid beamforming (HB) architecture, the optimal beamforming 
based 
on the fully TTD (FTTD) arrays, and the TTD-DPP and PDF methods built on the 
DPP 
architecture. The power consumption of these three architectures, denoted by 
$P_{\rm HB}$, $P_{\rm TTD}$, and $P_{\rm DPP}$ are given by~\cite{DPP_Tan2019}
\begin{align}
	P_{\rm HB} &= P_{\rm t} + 
	P_{\rm B} + N_{\rm RF}P_{\rm RF} + N_{\rm RF} N P_{\rm PS}, \notag\\
	P_{\rm FTTD} &= P_{\rm t} + 
	P_{\rm B} + N_{\rm RF}P_{\rm RF} + N_{\rm RF} N P_{\rm TTD}, \notag\\ 
	P_{\rm DPP} &= P_{\rm t} + 
	P_{\rm B} + N_{\rm RF}P_{\rm RF} + N_{\rm RF} N P_{\rm PS} + N_{\rm RF} K 
	P_{\rm TTD}, \notag 
\end{align}
where $P_{\rm t}$ and $P_{\rm B}$ represent the transmission power and digital 
processing power, and $P_{\rm RF}$, $P_{\rm PS}$, and $P_{\rm TTD}$ denote the 
power consumption of each RF chain, phase shifter, and TTD element.  The 
following 
typical values are adopted: $P_{\rm t}= 30\:{\rm mW}$~\cite{DPP_Tan2019}, 
$P_{\rm B} = 200\:{\rm 
mW}$~\cite{DPP_Tan2019}, 
$P_{\rm RF} = 250\:{\rm mW}$~\cite{HP_Mendez2016}, $P_{\rm PS} = 30\:{\rm 
mW}$~\cite{HP_Mendez2016}, 
and $P_{\rm TTD} = 
100\:{\rm mW}$~\cite{DPP_Tan2019}. The other settings are as follows: ${\rm 
SNR} = 5\:{\rm dB}$; 
$K = 8$; $N = 256$; $r_u\sim \mathcal{U}(1\:{\rm m}, 
30\:{\rm 
	m})$; $\theta_u \sim \mathcal{U}(-\frac{\pi}{3}, 
\frac{\pi}{3})$. 
It is clear from Fig.~\ref{img:EE_U} that 
even though the optimal beamforming has the 
highest spectral efficiency, its energy efficiency is pretty low because of the 
large number of high-power TTDs used, i.e., $UN$. In contrast, 
we can observe that the proposed PDF 
achieves much higher energy efficiency than all compared benchmarks. This 
observation is attributed to the fact that the PDF method can efficiently 
overcome the near-field beam-split effect with a quite small number of 
expensive TTDs, i.e., $UK \ll UN$, which further demonstrates the efficacy of 
the 
proposed PDF method.  

	\section{Conclusions}\label{sec:6}
	In this paper, we reveal an important challenge for future ELAA 
	communications, i.e., the near-field beam split effect.
	To address this challenge, we first propose a piecewise-far-field model to 
	approximate the near-field model with high accuracy. Applying this model, 
	a PDF method is proposed to efficiently alleviate the near-field beam 
	split effect through the joint manipulation of PSs and TDs. 
	Moreover, we define a new metric called as ``effective Rayleigh distance" 
	by evaluating the 
	beamforming gain, which is more accurate in quantifying the near-field 
	range than the classical Rayleigh distance 
	for practical communications. Finally, numerical results are provided to 
	demonstrate the effectiveness of our work.
	
	The discussion on the near-field beam split effect and our PDF method 
	provide new vision to ELAA beamforming. Besides, our proposed  
	effective Rayleigh distance offers a new way to evaluate the near-field 
	range. 
	For future works, people could investigate the near-field beam split effect 
	in more general situations, such as multi-antenna users, uniform planar 
	arrays, reconfigurable intelligent surfaces~\cite{XLRIS_Song2023}, and so 
	forth. In addition, 
	extending the effective Rayleigh distance to more applications  deserves 
	in-depth study as well.

\section*{Acknowledgment}
The authors sincerely thank Prof. Robert Schober and Prof. Lajos Hanzo for 
their 
valuable 
comments on this work. 

	\footnotesize
	\bibliographystyle{IEEEtran}
\bibliography{IEEEabrv,refs}

	\begin{IEEEbiography}[{\includegraphics[width=1in,height=1.25in,clip,keepaspectratio]{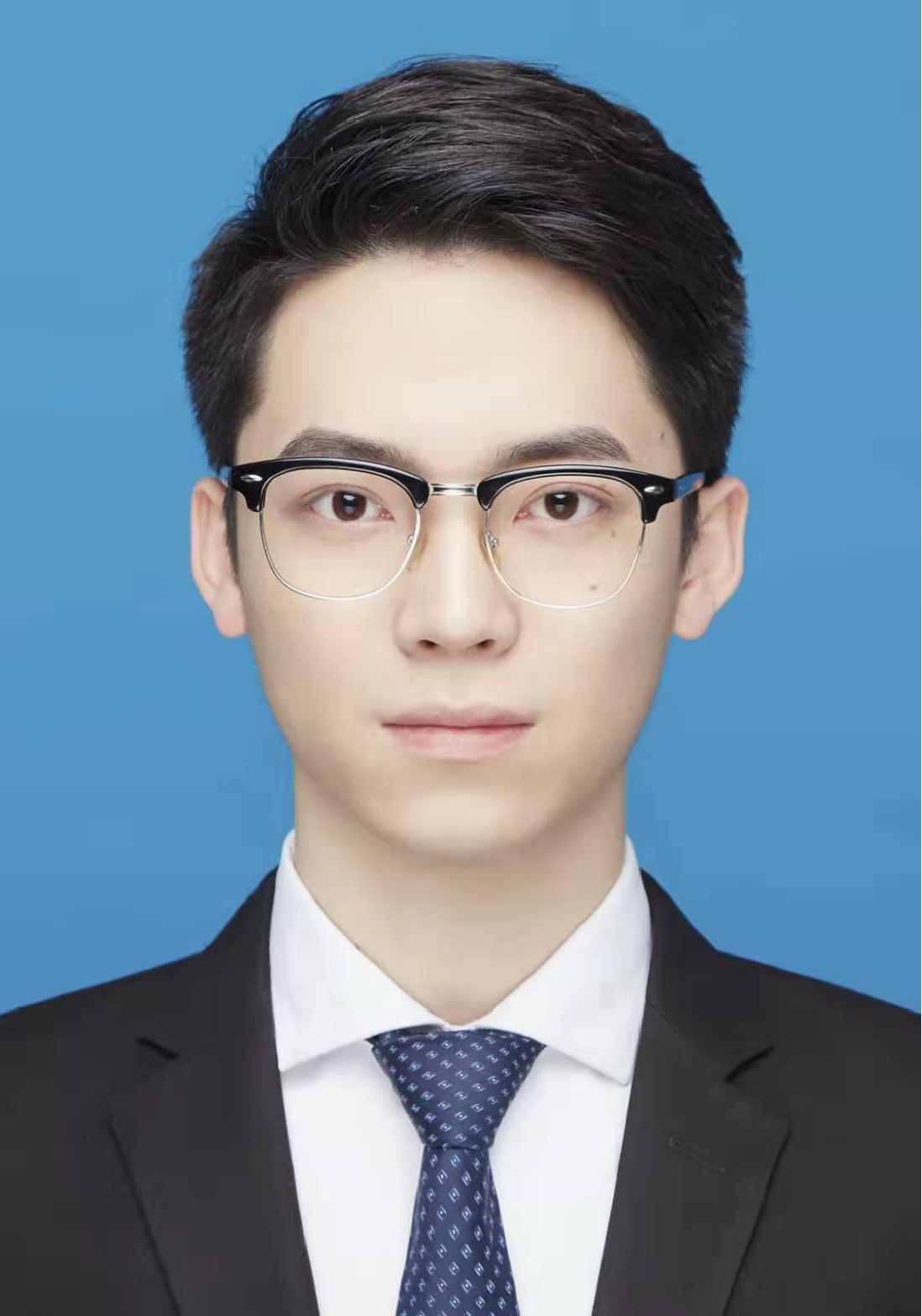}}]{Mingyao
	 Cui} received the
 B.E. and M.S. degrees in electronic engineering from Tsinghua University, 
 Beijing, China, in 2020 and 2023, respectively. He is currently pursuing the 
 Ph.D. degree in the University of HongKong. His research interests include 
 massive MIMO, millimeter-wave communications, and near
field communications. He received the IEEE ICC Outstanding 
Demo Award and the 
National Scholarship in 2022, and the HKPF Scholarship in 2023.
\end{IEEEbiography}

\begin{IEEEbiography}[{\includegraphics[width=1in,height=1.25in,clip,keepaspectratio]{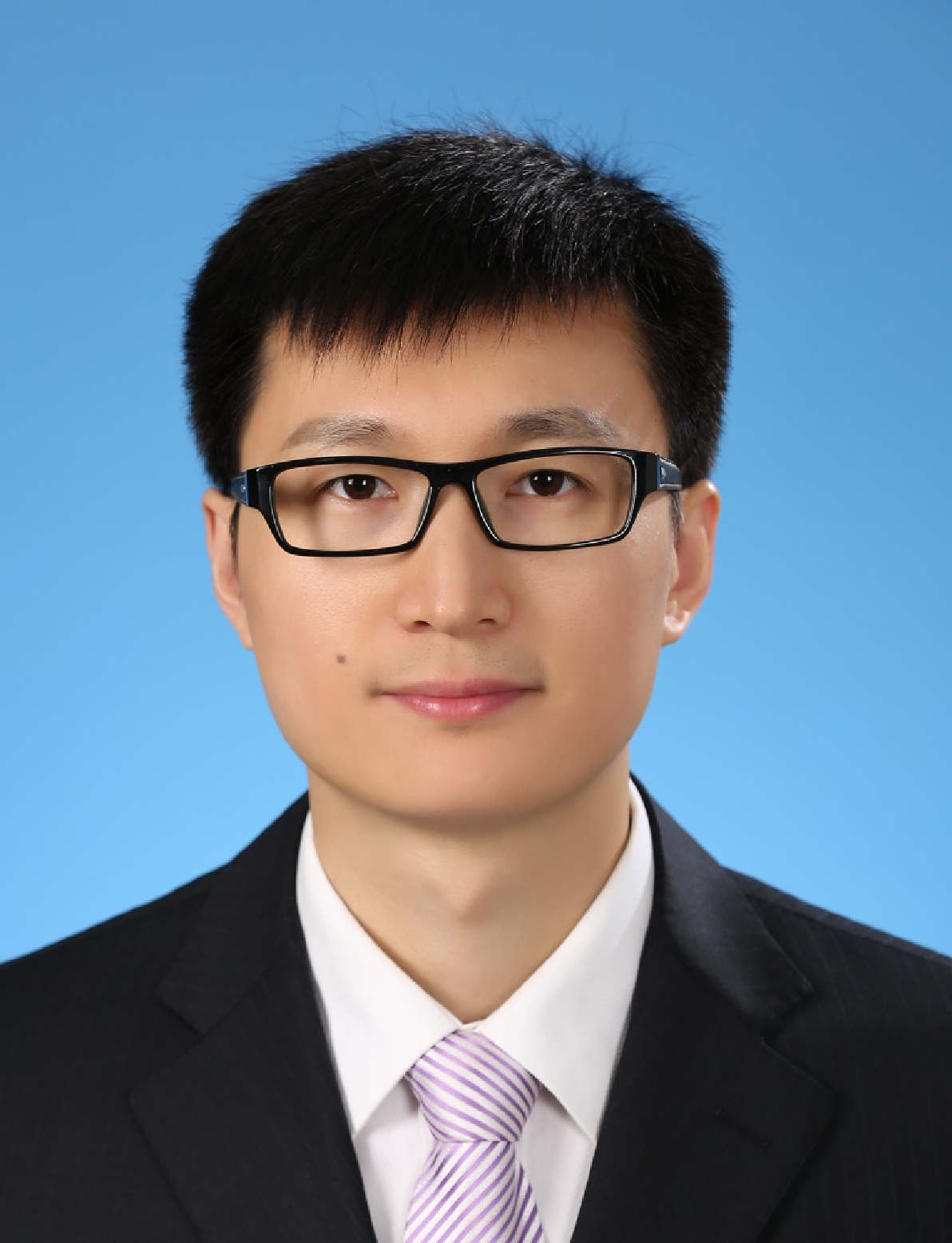}}]
{Linglong Dai} (Fellow, IEEE) received the B.S. degree from Zhejiang 
University, Hangzhou, China, in 2003, the M.S. degree (with the highest honor) 
from the China Academy of Telecommunications Technology, Beijing, China, in 
2006, and the Ph.D. degree (with the highest honor) from Tsinghua University, 
Beijing, China, in 2011. From 2011 to 2013, he was a Postdoctoral Research 
Fellow with the Department of Electronic Engineering, Tsinghua University, 
where he was an Assistant Professor from 2013 to 2016, an Associate Professor 
from 2016 to 2022, and has been a Professor since 2022. His current research 
interests include massive MIMO, reconfigurable intelligent surface (RIS), 
millimeter-wave and Terahertz communications, near-field communications, 
machine learning for wireless communications, and electromagnetic information 
theory.

He has coauthored the book {\it MmWave Massive MIMO: A Paradigm for 5G} 
(Academic Press, 2016). He has authored or coauthored over 100 IEEE journal 
papers and over 60 IEEE conference papers. He also holds over 20 granted 
patents. He has received five IEEE Best Paper Awards at the IEEE ICC 2013, the 
IEEE ICC 2014, the IEEE ICC 2017, the IEEE VTC 2017-Fall, the IEEE ICC 2018, 
and the IEEE GLOBECOM 2023. He has also received the Tsinghua University 
Outstanding Ph.D. Graduate Award in 2011, the Beijing Excellent Doctoral 
Dissertation Award in 2012, the China National Excellent Doctoral Dissertation 
Nomination Award in 2013, the URSI Young Scientist Award in 2014, the IEEE 
Transactions on Broadcasting Best Paper Award in 2015, the Electronics Letters 
Best Paper Award in 2016, the National Natural Science Foundation of China for 
Outstanding Young Scholars in 2017, the IEEE ComSoc Asia-Pacific Outstanding 
Young Researcher Award in 2017, the IEEE ComSoc Asia-Pacific Outstanding Paper 
Award in 2018, the China Communications Best Paper Award in 2019, the IEEE 
Access Best Multimedia Award in 2020, the IEEE Communications Society Leonard 
G. Abraham Prize in 2020, the IEEE ComSoc Stephen O. Rice Prize in 2022, the 
IEEE ICC Outstanding Demo Award in 2022, and the National Science Foundation 
for Distinguished Young Scholars in 2023. He was listed as a Highly Cited 
Researcher by Clarivate Analytics from 2020 to 2023. He was elevated as an IEEE 
Fellow in 2022.
\end{IEEEbiography}
	

\end{document}